\documentclass[prb,twocolumn,superscriptaddress,preprintnumbers,amsmath,amssymb,nofootinbib,floatfix,longbibliography]{revtex4-2}
\usepackage{graphicx,bm}
\usepackage{float}
\usepackage{subfigure}
\usepackage{amsmath}
\usepackage{cases}
\usepackage[overload]{empheq}
\usepackage{amsfonts}
\usepackage{amsthm, amssymb,bbm,bm}
\usepackage[utf8]{inputenc}
\usepackage[english]{babel}
\usepackage{xcolor}
\usepackage{bbold}
\usepackage{graphicx}
\usepackage{appendix}
\begin{document}
\title{Theory of exciton-polariton condensation in gap-confined eigenmodes}
\author{Davide Nigro}
\affiliation{Dipartimento di Fisica, Universit\`{a} di Pavia, via Bassi 6, I-27100 Pavia, Italy}
\author{Dario Gerace}
\affiliation{Dipartimento di Fisica, Universit\`{a} di Pavia, via Bassi 6, I-27100 Pavia, Italy}

\begin{abstract}
Exciton-polaritons are bosonic-like elementary excitations in semiconductors, which have been recently shown {to display} large occupancy of topologically protected polariton bound states in the continuum in suitably engineered photonic lattices [Nature {\bf 605}, 447 (2022)], compatible with the definition of polariton condensation. However, a full theoretical description of such condensation mechanism that is based on a non equilibrium Gross-Pitaevskii formulation is still missing. Given that the latter is well known to account for polariton condensation in conventional semiconductor microcavities, here we report on its multi-mode generalization, showing that it allows to fully interpret the recent experimental findings in patterned photonic lattices, including emission characteristics and condensation thresholds. Beyond that, it is shown that the polariton condensation in these systems is actually the result of an interplay between negative mass confinement of polariton eigenstates (e.g., due to the photonic gap originated from the periodic pattern in plane) and polariton losses. We are then able to show that polariton condensation can also occur in gap-confined  bright modes, i.e., coupling of QW excitons to a dark photonic mode is not necessarily required to achieve a macroscopic occupation with low population threshold. 
\end{abstract}
\maketitle

\section{Introduction}
Exciton-polaritons arising in low-dimensional nanostructures, such as coupled quantum well (QW) excitons and confined photonic modes, have been shown to behave as a weakly interacting Bose gas \cite{Sanvitto2016}. Condensation phenomena in these systems have to be inevitably described in terms of a balance between driving and losses, due to photon leakage or exciton recombination. In this respect, a suitable generalization of the standard nonlinear Schr\"{o}dinger equation, known as the non-equilibrium Gross-Pitaevskii equation (NEGPE), is able to largely account for the observed phenomenology including external driving and intrinsic losses, either under resonant or non-resonant pumping of the exciton-polariton field \cite{Carusotto_Ciuti_RMP2013}. In particular, spectacular phenomena such as Bose-Einstein condensation \cite{Kasprzak2006,Balili2007}, superfluidity \cite{Amo2009}, formation of quantized vortices \cite{Lagoudakis_nphys_2008} have been observed in planar semiconductor microcavities, in which the photon field is confined in the QW plane between two high-reflectivity Bragg mirrors \cite{Kavokin_Microcavities_book}. Besides showing interest for fundamental physics studies, exciton-polaritons have also been shown to potentially allow for useful applications due to their superior nonlinear properties, such as low-threshold lasers \cite{Azzini2011}, all-optical switching \cite{Ballarini2013}, sensing \cite{Paschos2018sensing}, etc.

\begin{figure}[b]
    \centering
    \includegraphics[scale=0.5]{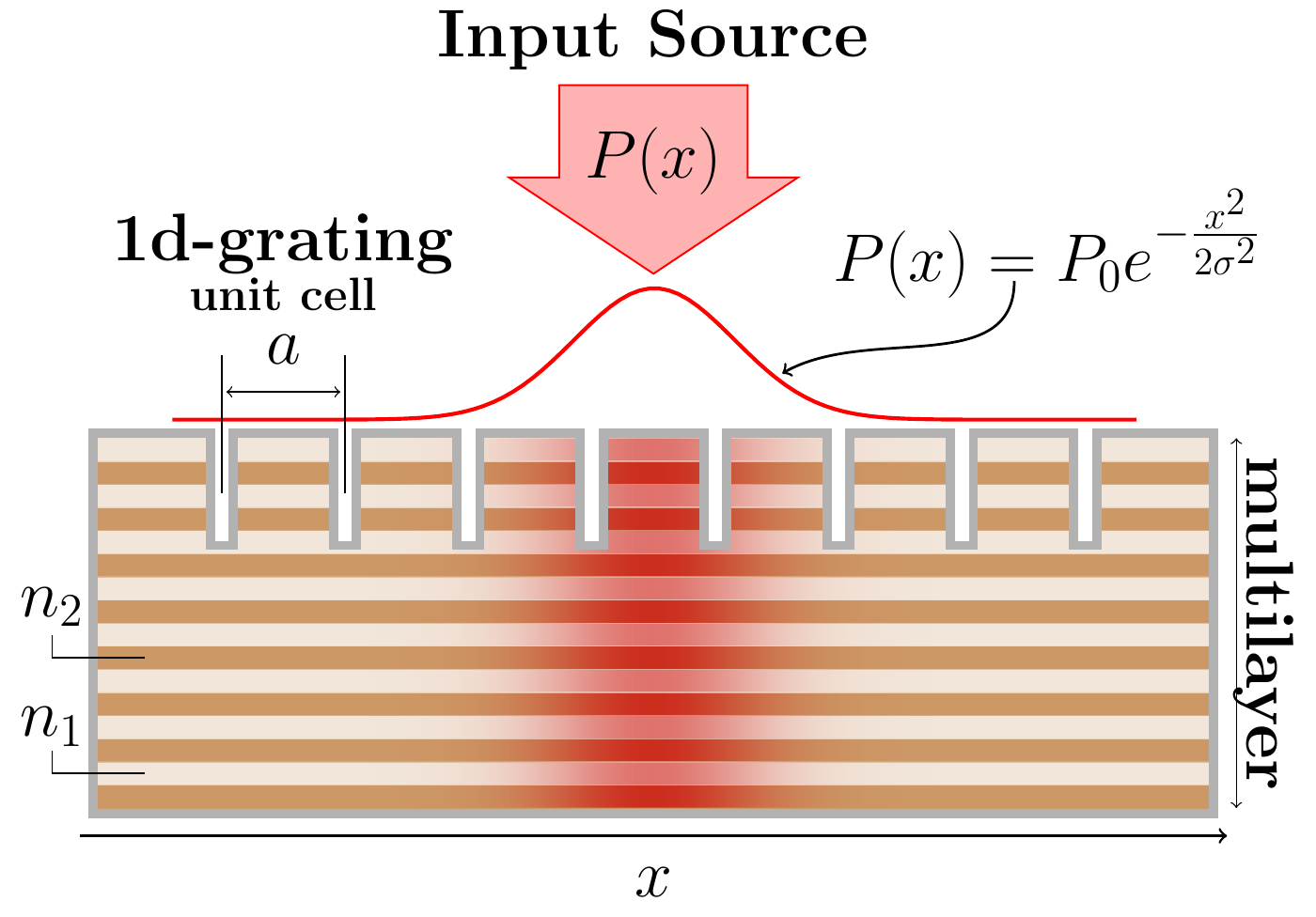}
    \caption{Schematic representation of the multi-QW heterostructure with surface periodic patterning that creates a photonic gap for in-plane propagating modes. Elementary excitations can be created by a laser spot $P(x)$, which locally changes the refractive index in the layers underneath ($n_1$ and $n_2$), thus inducing an effective confining potential for the negative mass polaritons. }
    \label{fig:sketch_platform}
\end{figure}

Recently, polariton condensation has been observed also in multi-layered QW heterostructures
with periodic surface patterning, such as the one sketched in Fig.~\ref{fig:sketch_platform}. In fact, it has been shown that top/bottom mirrors are not necessary to achieve the polariton condensation threshold, since out-of plane radiation losses associated to the photonic component of the polariton field can be fully engineered by the photonic lattice \cite{GME2006}, even a one-dimensional one \cite{GeracePRE2004}. In particular, radiative losses of the photonic component can be fully suppressed by symmetry along certain directions, leading to the currently widespread concept of photonic bound states in the continuum (BIC) \cite{Hsu2016,Azzam2021}. When coupled to QW excitons, these modes lead to the formation of polariton eigenmodes that are dark for emission along specific spatial directions and at fixed energy \cite{KoshelevPRB2018}, inheriting the topologically protected nature of the purely photonic BIC \cite{Zanotti2022,HaMyDang_AOM2022,Kim_NanoLett_2021}. Polariton condensation has been observed in these systems by efficient relaxation into modes arising from the saddle-like dispersion of the lower polariton branch, i.e., not an absolute minimum of the dispersion relation \cite{Ardizzone_2022,Riminucci22}. In these works, the condensation mechanism has been qualitatively interpreted as due to the creation of an effective potential well for negative mass polaritons, induced by the exciting laser spot over a finite width along the periodicity lattice, as well as to the fact that such confined states acquire a longer lifetime than other modes due to coupling to purely photonic BICs. However, no quantitative theory has been provided to account for the condensation in such BIC-like modes, to date. In particular, a rigorous discussion about the role of the negative mass branch in connection with the photonic-induced gap obtained as a consequence of the periodic refractive index modulation is still missing. This manuscript aims at filling this theoretical gap, by developing a model of out-of-equilibrium condensation in a multi-band exciton-photon coupled system with symmetry-dependent losses. In particular, we hereby unravel the critical role played by the negative mass and the energy gap between dark and bright modes in reaching the condensation onset.

Here is a short outline of the results presented in the paper. In Sec.~\ref{sec:theoretical_model} we describe the dynamical equations used to picture the onset of polariton condensation in the continuous-wave (CW) regime of a model Hamiltonian with a given number of photonic branches coupled to QW exciton modes. Then, in Sec.~\ref{sec:two-modes} we address in detail the behavior of our effective theory in a specific case, where the relevant physics can be traced back to the presence of two counter-propagating photonic modes. In order to understand the polariton condensation in this theoretical framework, we first characterize the polariton dispersion arising from the Hamiltonian diagonalization, by deriving polariton bands in Sec. \ref{sec:polariton_bands}, and the effects of an external space-dependent potential $V(x)$ coupled to the particle density in Sec. \ref{sec:gap_states}. In Sec. \ref{sec:pol_cond_two} we show the results concerning the polariton condensation in discrete energy eigenvalues  lying within the energy gap, which is ultimately due to the periodic photonic modulation. These eigenvalues correspond to spatially confined eigenmodes below the potential $V(x)$. In particular, after showing the results of our numerical simulations describing the behavior of the condensate population in the steady-state, we provide an analysis of the condensate loss rate (Sec.~\ref{sec:anal_linewidth}), as well as the spectral density describing the energy-momentum behavior of the light emitted by the condensate (Sec.~\ref{sec:light_emission}). The behavior of such quantities in all the cases considered in the present work evidently supports the conjecture that polaritons are condensing into the discrete gap-confined eigenmodes previously characterized in Sec.~\ref{sec:gap_states}. Depending on the dark or bright nature of the negative mass branch out of which they originate, these modes display very different direct and reciprocal space profiles. In light of these results, in Sec.
\ref{sec:summary} we finally discuss the relevance of our findings and future perspectives.

\section{Theoretical model}\label{sec:theoretical_model}
In this section we introduce a general dynamical model that can be used to describe and characterize the onset of polariton condensation and its relation to the eigenstates properties in heterostructures like the one sketched in Fig. \ref{fig:sketch_platform}. We will assume  CW pumping in this work, although the model is general. The system is assumed to be uniform in the direction transverse to $x$, consisting of several layers with different refractive indices, in the figure denoted as $n_1$ and $n_2$. For instance \cite{Ardizzone_2022,Riminucci22}, such  a stack of different index material might correspond to a set of GaAs/Al$_x$Ga$_{1-x}$As layers, which give rise to the formation of quantum well (QW) excitons in the lower band-gap material \cite{Andreani1990,Savona1995}. 
Due to the top surface patterning with spatial periodicity $a$, the free-propagating electromagnetic modes get folded, leading to the formation of gapped Bloch resonances \cite{GME2006,GeracePRE2004}. When coupled to QW excitons, these photonic modes are shown to give rise to the concept of photonic crystal polaritons \cite{GeracePRB2007}, which have been evidenced in different systems and material platforms \cite{BajoniPRB2009,Deng2018_ncomm,HaMy2020,Majumdar2020}, all of them falling within the same theoretical treatment \cite{Zanotti2022}. In particular, it has been evidenced that a surface patterning is sufficient to induce 
gapped polariton branches with positive and negative effective masses around normal incidence and energies below the bare exciton resonance \cite{Ardizzone_2022,Riminucci22}. It has been observed that polariton condensation occurs within a set of discretized quantum states appearing  within such energy gaps. Their emission properties are shown to depend both on the spatial profile of the input light-source \cite{Ardizzone_2022,Gianfrate2023}, $P(x)$ in Fig.~\ref{fig:sketch_platform}, and on the intrinsic emission features of the lower-lying polariton branch delimiting the band-gap. 
Similarly to the description of polariton condensation in the planar microcavity case, here it is reasonable to assume that particle scattering towards the condensed eigenstate can be described by means of an exciton reservoir non-linearly coupled to polariton states \cite{Wouters_Carusotto_PRL,Baboux2018,Fontaine_Nature_KPZ,Amelio2023_kPZ_arxiv}. However, in the present case it is also crucial to account for the mechanism leading to the formation of such discrete states within the polariton gap, as well as the peculiar emission properties shown by either the bare polariton bands or the polariton condensate modes. Moreover, the presence of different photonic modes may lead to some of them being in weak coupling with the exciton, but still participating in the relaxation dynamics and thus worth being taken into account. In this respect, as it is shown in the following, our approach seems to provide a quite complete picture allowing to account for the observed phenomenology, as well as predicting possible new outcomes, when including all these crucial aspects. 

In our model, similarly to the standard approach adopted to describe  condensation in microcavity systems \cite{Wouters_Carusotto_PRL,Wouters_Carusotto_Ciuti_PRB,Lagoudakis_Inactive_condensate}, we assume $P(x)$ to be incoherently coupled to a reservoir population, $n(x,\,t)$, describing high-energy exciton states. By stimulated scattering, such a reservoir feeds lower-energy eigenmodes, thus replenishing the exciton-polariton population in these states. However, in order to capture the complex band-gap physics, emission properties as well as the formation of discrete levels, instead of using a single component wavefunction $\psi(x,\,t)$ (usually describing the lower branch polariton field), here we assume that the reservoir density couples to a set of relevant bare photonic and excitonic modes of the structure. In other words, the system dynamics is hereby described by means of a multi-component vector, $\overrightarrow{\psi}(x,\,t)$ formally reading
\begin{equation}\label{eq:vectorial_psi}
\overrightarrow{\psi}(x,\,t)=\left(
A_1,A_2,\cdots,A_{L},X_1,X_2,\cdots,X_L\right)^{T},
\end{equation}
in which $A_l\equiv A_{l}(x,\,t)$ and $X_l\equiv X_{l}(x,t)$ ($l\in [1,2,\,\cdots,\,L]$) denote the wavefunction of the $l$-th photonic and $l$-th excitonic mode respectively. In particular, $2L$ is the total number of polariton branches participating in the system dynamical evolution.\\
By setting $n\equiv n(x,\,t)$, $\overrightarrow{\psi}\equiv\overrightarrow{\psi}(x,\,t)$, we assume the reservoir-polariton dynamics to be described by the following set of coupled differential equations
\begin{subequations}
\begin{align}[left ={ \,\empheqlbrace}]
& \frac{d}{dt}n =P(x)-\frac{1}{\tau_R}\,n- g n  \langle\overrightarrow{\psi},\,G\overrightarrow{\psi}\rangle \label{eq:firstline} \\
&\frac{d}{dt}\overrightarrow{\psi}= \tilde{H}(x)\overrightarrow{\psi}+\frac{g}{2}n\, G\,\overrightarrow{\psi}\label{eq:secondline}
\end{align}
\label{eq:dynamical_system}
\end{subequations}
in which $\langle \vec{w},\,\vec{z}\rangle=\sum_{j}w_j^*z_j$ denotes the hermitian scalar product between the two vectors $\vec{w}$ and $\vec{z}$, with $w_j^*$ denoting the complex-conjugated of the component $w_j$. 

The different terms in Eq.~\eqref{eq:firstline} account for three main physical contributions: (i) the injection of population due to the CW driving through $P(x)$; (ii) the presence of a loss term controlled by the intrinsic finite-lifetime $\tau_R$ of the reservoir; (iii) a non-linear decay term proportional to the effective coupling $g>0$, that describes the stimulated population scattering from the reservoir to photon and exciton states. The $2L\times 2L$ matrix $G$ represents an operator satisfying the following constraints:
\begin{subequations}
\begin{align}[left ={ \,\empheqlbrace}]
& G^{\dagger}=G &\mbox{(hermiticity)} \label{eq:G-hermiticity}\\
& G\geq 0 & \mbox{(positive semi-definite)}\label{eq:G-constraint-positivity}\\
& \sum_{j} G_{jj}=1 & \mbox{(unit trace)} \label{eq:G-constraint-normalization}
\end{align}
\label{eq:G-constraint}
\end{subequations}
Equations \eqref{eq:G-hermiticity} and \eqref{eq:G-constraint-positivity} ensure that the reservoir density evolves as a real quantity, with last term acting as an effective loss (i.e., gain on the  $\vec{\psi}$ field, as clear from Eq.~\eqref{eq:secondline}). The normalization constraint in Eq.~\eqref{eq:G-constraint-normalization} guarantees for a particle-conserving dynamics. Indeed, if on the one hand $g$ controls the global scattering rate from the reservoir, on the other the model accounts for several decay channels, each controlled by an element of the matrix $G$. Since a particle scattered away from the reservoir must create a single particle into the light-matter part of the model, the $\{G_{jj}\}$ must add up to 1.

The coupled exciton-photon dynamics is encoded into Eq.~\eqref{eq:secondline}. The term proportional to the (non-Hermitian) $2L\times 2L$ matrix operator $\tilde{H}(x)$, given by
\begin{equation}\label{eq:nonhermitian_model}
\tilde{H}(x)= -i \frac{E(x)}{\hbar}-\Gamma(x)-i\frac{W(x)}{\hbar}
\end{equation}
accounts for the peculiar complex energy-momentum dispersion laws of the relevant bare modes and their mutual linear interaction, i.e., $-i E(x)/\hbar -\Gamma(x)$, as well as the presence of an external space-dependent potential, $W(x)$, such that $W(x)=W^{\dagger}(x)$. In particular, for the same reason leading to Eq.~\eqref{eq:G-constraint}, in order to interpret $\Gamma(x)$ as a proper decay operator, one must require $\Gamma(x)$ to satisfy the following constraints:
\begin{subequations}
\begin{align}[left ={ \,\empheqlbrace}]
& \Gamma^{\dagger}(x)=\Gamma(x) &\mbox{(hermiticity)} \label{eq:Gamma-hermiticity}\\
& \Gamma(x)\geq 0 & \mbox{(positive semi-definite)}\label{eq:Gamma-constraint-positivity}
\end{align}
\label{eq:Gamma-constraint}
\end{subequations}
Finally, the last term in Eq.~\eqref{eq:secondline} accounts for the particle transfer from the reservoir to photonic and excitonic modes. In particular, the structure of the nonlinear terms proportional to $g$ in Eq.~\eqref{eq:dynamical_system}, and the constraints imposed on $G$, lead to the expected $g$-independent behavior of the time derivative of the total particle density, $N(x,\,t)\equiv n(x,\,t)+ \langle \overrightarrow{\psi}(x,\,t),\,\overrightarrow{\psi}(x,\,t)\rangle$. 
By combining the two expressions in Eq.~\eqref{eq:dynamical_system}, the following rate equation for $N(x,t)$ is obtained
\begin{equation}\label{eq:total_density_evolution}
\begin{split}
&\frac{d}{dt}N(x,\,t)=\frac{d}{dt}n(x,\,t)+\frac{d}{dt}\langle \overrightarrow{\psi}(x,\,t),\,\overrightarrow{\psi}(x,\,t)\rangle=\\
&=P(x)-\frac{1}{\tau_R}\,n-\langle \tilde{H}(x)\overrightarrow{\psi},\,\overrightarrow{\psi}\rangle+\langle \overrightarrow{\psi},\,\tilde{H}(x)\overrightarrow{\psi}\rangle.\\
\end{split}
\end{equation}
In the long-time limit, since the CW source does not depend on time (i.e., the only driven Fourier component of the reservoir density is the $\omega=0$ one), it is reasonable to assume that the dynamical system does not produce oscillating solutions (limit cycles). As a consequence, the steady-state (ss) configuration, i.e., the solutions satisfying $\frac{d}{dt}n(x,\,t)=\frac{d}{dt}\overrightarrow{\psi}(x,\,t)=0$,  correspond to a fixed point of the dynamical system in Eq.~\eqref{eq:dynamical_system}. In other words, by recalling that due to Eq.~\eqref{eq:G-constraint-positivity} $\langle \vec{\psi},G\vec{\psi}\rangle \geq 0\, \forall\,\vec{\psi}$, we have the steady state relations 
\begin{subequations}
\begin{align}[left ={ \,\empheqlbrace}]
& n_{ss}(x)= \frac{P(x)}{\frac{1}{\tau_R}+ g \langle\overrightarrow{\psi}_{ss}(x),\,G\overrightarrow{\psi}_{ss}(x)\rangle} \label{eq:first_eq_fp} \\
&\tilde{H}(x)\overrightarrow{\psi}_{ss}(x)=-\frac{g}{2}\frac{P(x)G\,\overrightarrow{\psi}_{ss}(x)}{\frac{1}{\tau_R}+ g \langle\overrightarrow{\psi}_{ss}(x),\,G\overrightarrow{\psi}_{ss}(x)\rangle}\, \label{eq:second_eq_fp}
\end{align}
\label{eq:fixed_point_solution}
\end{subequations}
in which $n_{ss}$ and $\overrightarrow{\psi}_{ss}$ denote the steady-state reservoir density and exciton-photon configuration, respectively. As it is easy to verify by direct inspection, the system of coupled equations  \eqref{eq:fixed_point_solution} always admits the trivial configuration corresponding to an empty exciton-photon subsystem and a reservoir density proportional to the CW profile as a fixed point solution, i.e., 
\begin{equation}\label{eq:trivial_fp}
    \overrightarrow{\psi}_{ss}=0,\quad n_{ss}(x)=\tau_R P(x).
\end{equation}
Depending on the pump strength $P_0$ (see Fig.~\ref{fig:sketch_platform}), we notice that such a 
solution might become linearly unstable, i.e., a small perturbation $\delta  \overrightarrow{\psi}(x)$ placed on top of $ \overrightarrow{\psi}_{ss}=0$ gives rise to a macroscopic increasing of the condensate density, i.e., $\langle \delta \overrightarrow{\psi}(x),\delta \overrightarrow{\psi}(x)\rangle$, pushing the system away from the trivial configuration in Eq.~\eqref{eq:trivial_fp}. By linearizing the dynamical system in Eq.~\eqref{eq:dynamical_system}, we obtain the following expression for the exciton-photon density evolution:
\begin{equation}
\begin{split}
    \frac{d}{dt}\langle \delta \overrightarrow{\psi}(x),\delta \overrightarrow{\psi}(x)\rangle&=-2\langle \delta \overrightarrow{\psi}(x),\Gamma(x)\delta\overrightarrow{\psi}(x)\rangle\\
    &+g\tau_R P(x)\langle \delta \overrightarrow{\psi}(x),G\delta  \overrightarrow{\psi}(x)\rangle,
\end{split}
\end{equation}
which implies that the empty solution becomes locally unstable whenever 
\begin{equation}
    \frac{g\tau_R}{2} P(x)G \geq \Gamma(x),
\end{equation}
that is when the particle gain induced by the CW source overcomes particle losses.

We conclude this section with a final remark. Even though we have reported a general formalism to account for $L$ photonic modes coupled to $L$ exciton states, in what follows we will restrict our attention to the $L=2$ case. Indeed, if on the one hand all the expressions previously reported hold true in the general case, on the one other hand such case study is of particular interest in light of the recent experiments \cite{Ardizzone_2022,Riminucci22,Gianfrate2023}, where it has been observed that the relevant dynamics leading to the onset of condensation only involves a pair of polariton bands below the exciton energy.
\section{Two photonic modes: spectral analysis}\label{sec:two-modes}
As discussed in Sec.~\ref{sec:theoretical_model}, in order to describe the onset of condensation we first address the spectral properties of the system. In particular, since we can safely assume the QW exciton energy-momentum dispersion to be flat, the only non-trivial part concerns the representation of the counter-propagating gapped Bloch resonances, in which the gap is induced by the periodic dielectric modulation, e.g., as sketched of Fig. \ref{fig:sketch_platform}. Even though a complete characterization of such modes in a one-dimensional (1D) periodic dielectric structure can be performed through a Maxwell solver based, e.g., on a guided-mode expansion  \cite{GeracePRE2004,GME2006}, here we rather follow the simplified approach discussed in \cite{Lu2020}, where it was shown that close to normal incidence (i.e., around $k=0$), the dispersion of gapped photonic branches can be reproduced and approximated by means of two counter-propagating modes with linear dispersion that are diffractively coupled by an off-diagonal term $U$. The latter can be either positive or negative, which is actually related to the composition of the unit cell of the periodic 1D lattice (see, e.g., Ref.~\cite{Ardizzone_2022}). In fact, we assume it as an effective model parameter, but we keep in mind that it can be fully engineered through the characteristics of the periodic pattern. We then consider the following matrices to build Eq.~\ref{eq:nonhermitian_model}
\begin{equation}\label{eq:E_two}
E(x)=\hbar
\left(
\begin{array}{cccc}
\omega_A -i  v_g\partial_x & U & \Omega_R & 0 \\
U & \omega_A +i  v_g\partial_x & 0 & \Omega_R \\
\Omega_R & 0 & \omega_X & 0 \\
0 & \Omega_R & 0 & \omega_X \\
\end{array}\right),
\end{equation}
\begin{equation}\label{eq:Gamma_two}
\Gamma(x)=\left(
\begin{array}{cccc}
\gamma_A & \tilde{\gamma}_A & 0 & 0\\
\tilde{\gamma}_A & \gamma_A & 0 & 0\\
0& 0 &\gamma_X & 0\\
0& 0 &0 & \gamma_X\\
\end{array}
\right)\equiv \Gamma_0,
\end{equation}
and
\begin{equation}\label{eq:V_two}
W(x)= V(x) \mathbb{1}_4,
\end{equation}
with $\mathbb{1}_4$ being the $4\times 4$ identity operator. Notice that we hereby assume that the potential $V$ is the same for either photonic or excitonic components, for simplicity of computations and without loss of generality. In fact, while in practical cases one might differentiate between photonic and excitonic contributions, the qualitative behavior of results we are going to discuss do not depend from the choice made here. In this scenario, the wavefunction $\overrightarrow{\psi}(x,\,t)$ reduces to the following four-components vector:
\begin{equation}\label{eq:four_com_psi_x}
\overrightarrow{\psi}(x,\,t)=\left( A_+(x,t),A_-(x,t),X_+(x,t),X_-(x,t)\right)^{T}
\end{equation}
This model describes two photonic modes separately coupled to two degenerate excitonic states, in the presence of an external potential $V(x)$ that is directly related to the presence of the CW source; $A_+$ and $A_-$ describe counter-propagating bands with linear dispersion whose slope is directly defined by their group-velocity, $v_g$. These modes cross at $k=0$ with energy $\hbar \omega_A$, and they have an intrinsic loss rate given by $\gamma_A$. The two modes $X_+$ and $X_-$ are then associated to a flat QW exciton resonance at energy $\hbar \omega_X$, whose bare loss rate is $\gamma_X$. The photonic modes are assumed to be linearly coupled via a term proportional to $\hbar U- i\hbar \tilde{\gamma}_A$, which converts right ($+$) propagating photons into left ($-$) propagating ones, and viceversa. In particular, in order to ensure that $\Gamma_0$ describes a proper decay term, one must require that $\gamma_A\geq \vert\tilde{\gamma}_A\vert$ (see Eq. \eqref{eq:Gamma-constraint}). Each photonic component is assumed to be independently coupled to an exciton mode, with a coupling rate set by the Rabi frequency $\Omega_R/(2\pi)$.

 As it is shown in the following, such a minimal model is already able to account for the onset of polariton condensation into discrete levels appearing inside the energy-gap between polariton bands with opposite effective-mass. In close agreement with experimental results reported in the literature, the properties of such levels is shown to depend on both the bare polariton dispersion and the pump characteristics. We first pay attention to the spectral properties of the non-Hermitian operator  $\tilde{H}(x)$. For the sake of clarity, we proceed in two steps. In Sec.~\ref{sec:polariton_bands}, we first characterise the polariton dispersion in the absence of the external potential $V(x)$. The role of such space-dependent term in the appearence of a set of discrete levels is addressed in Sec.~\ref{sec:gap_states}.
 
\subsection{Eigenmodes: complex Polariton dispersion for  $W(x)=0$}\label{sec:polariton_bands}
The complex eigenmodes of the exciton-photon coupled system can be obtained as real and imaginary polariton bands by diagonalization of the operator  
\begin{equation}\label{eq:model_no_potential}
\tilde{H}_0(x)=-i \frac{E(x)}{\hbar} -\Gamma_0,
\end{equation}
where $E(x)$ and $\Gamma_0$ are given in Eq.s~\eqref{eq:E_two} and \eqref{eq:Gamma_two}, respectively. To this purpose, let us consider a generic four-component wavefunction $\overrightarrow{\psi}(x,t)$, as the one in Eq.~\ref{eq:four_com_psi_x}, and rewrite it in terms of its Fourier components with respect to the real space coordinate $x$. By applying the operator \eqref{eq:model_no_potential} formally gives
\begin{equation}
\tilde{H}_0(x)\overrightarrow{\psi}(x,\,t)= \int\frac{\mbox{d}k}{2\pi} e^{ikx} \tilde{H}_0(k)\overrightarrow{\psi}(k,\,t),
\end{equation}
with $\tilde{H}_0(k)\equiv-i \frac{E_0(k)}{\hbar}-\Gamma_0$, where
\begin{equation}\label{eq:E_k_two}
E_0(k)=\hbar
\left(
\begin{array}{cccc}
\omega_A +  v_g\,k & U & \Omega_R & 0 \\
U & \omega_A -  v_g\,k & 0 & \Omega_R \\
\Omega_R & 0 & \omega_X & 0 \\
0 & \Omega_R & 0 & \omega_X \\
\end{array}\right).
\end{equation}
Since $\tilde{H}_0(k)$ satisfies the conditions $
\tilde{H}^{\dagger}_0(k)\tilde{H}_0(k)=\tilde{H}_0(k)\tilde{H}^{\dagger}_0(k)$, by spectral theorem it can be diagonalized by means of a unitary operator, $U_k$. In the present case the diagonalization can be performed analytically. The resulting polariton eigenmodes (complex eigenvalues) are explicitly reported in App.~\ref{app:pol_bands}.

\begin{figure}[t]
\centering
\includegraphics[scale=0.45]{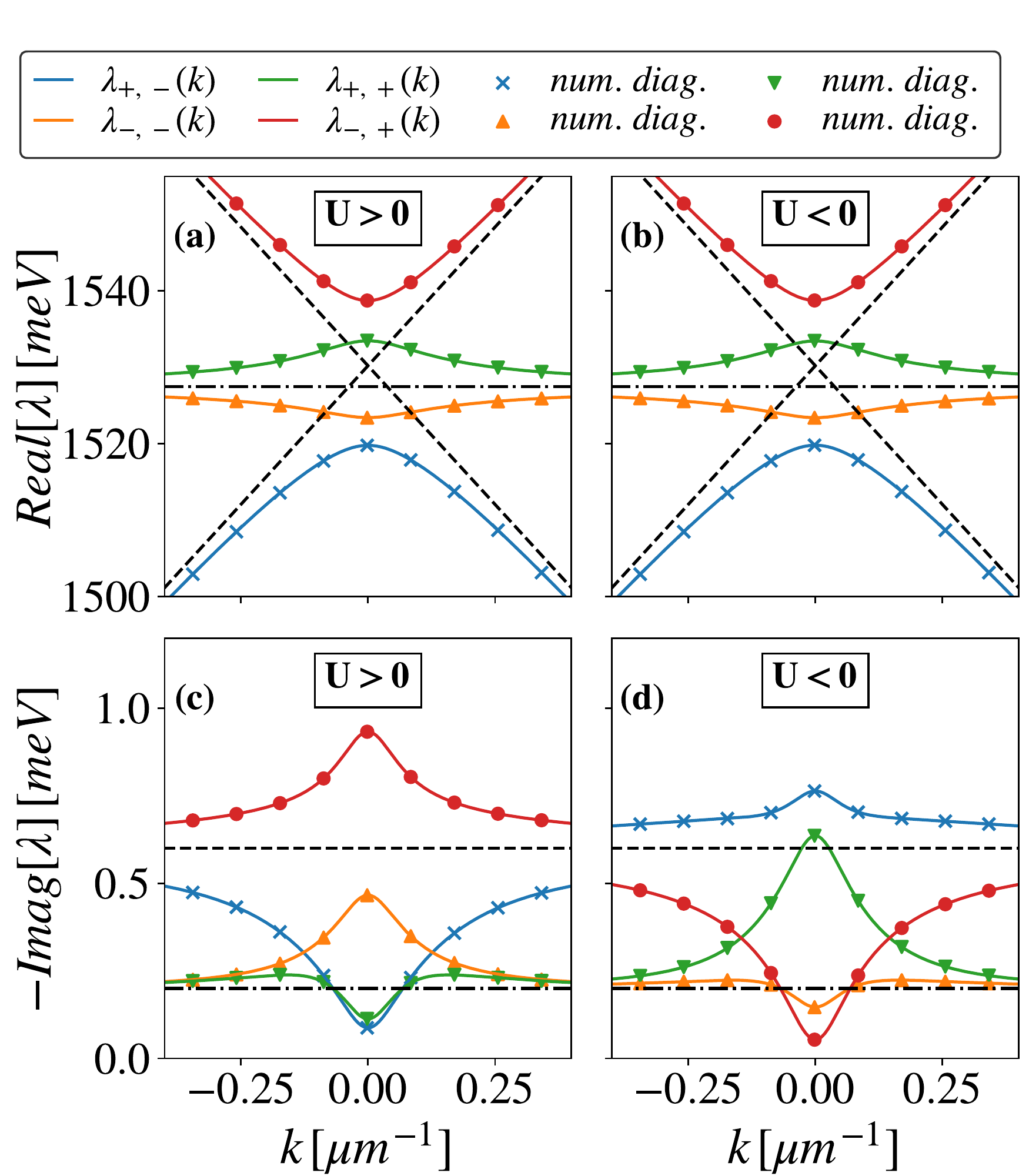}
\caption{Energy-momentum dispersion of the four polariton complex eigenmodes obtained from the model in Eq.~\eqref{eq:model_no_potential}. First, the real part of the polariton dispersion ($Real[\lambda]$) is shown for (a) $\hbar U=4.45$ meV and (b) $\hbar U=-4.45$ meV, respectively. The negative of the corresponding imaginary part  ($-Imag[\lambda]$) is shown for (c) the positive and (d) negative $U$ cases of (a) and (b), respectively. In all the panels, solid lines correspond to the analytical expressions reported in Eqs.~\eqref{eq:polariton_plus} and \eqref{eq:polariton_minus}. Points labeled with ``num. diag." correspond to the results of the numerical diagonalization. Dashed lines in panels (a), (b),(c) and (d) describe the properties of the bare photonic dispersions, i.e., modes $A_+(k)$ and $A_-(k)$. Dot-dashed lines describe the two degenerate exciton bands. The other relevant parameters are set as follows: $\hbar \omega_X= 1527.5$ meV; $\hbar\gamma_X=0.2$ meV; $\hbar \omega_A= 1530$ meV; $\hbar \gamma_A=\hbar\tilde{\gamma}_A=0.6$ meV; $v_g=110$ $\mu$m/ps and $\hbar=0.66$ meV$\cdot$ps. }\label{fig:pol_bands_example}
\end{figure}

Numerical diagonalization leads to the results shown in Fig.~\ref{fig:pol_bands_example}, where perfect matching is obtained for the eigenvalues of $i\hbar\tilde{H}_0(k)$ obtained numerically and the analytical expressions $\lambda_{\alpha,\,\beta}(k)$ given in Eqs.~\eqref{eq:polariton_plus} and \eqref{eq:polariton_minus}. In particular, the parameters used in the present case are compatible with typical values reported for inorganic semiconductor samples~\cite{Riminucci22}. As expected, at large $\vert k \vert$ the polariton eigenvalues asymptotically approach the bare photonic and excitonic ones, both in terms of real and imaginary parts. Also, and at difference with  microcavity polariton dispersions \cite{Sanvitto2016}, this polariton band strucutre displays a number of very peculiar features close to exciton-photon resonance. Independently of the sign of $U$, the bands $\lambda_{+,-}$ and $\lambda_{+,+}$ are characterized by an effective negative-mass $m^*$ (where we define $m^{*}=\partial_k^2 \lambda(k)\vert_{k=0}<0$), while $\lambda_{-,+}$ and $\lambda_{-,-}$ are both characterized by a positive one, as evident from Fig.~\ref{fig:pol_bands_example}(a,b). In both cases, bands having opposite effective mass are separated by an energy gap. In addition, in the vicinity of $k=0$ the bands are characterized by a $k$-dependent imaginary part (i.e., loss rate). As shown in Fig. \ref{fig:pol_bands_example}(c), for $U>0$ intrinsic losses of negative-mass polaritons are smaller than the exciton one, while positive-mass polariton loss rates are clearly larger than $\hbar\gamma_X$. As shown in Fig.~\ref{fig:pol_bands_example}(d), the scenario is reversed for negative values of $U$, where we see that positive-mass polaritons are characterized by small losses, when compared to the exciton states. Such a behavior is compatible with the one reported in Ref.~\cite{Lu2020}, where a band inversion between a bright and a dark mode is observed upon inverting the sign of the diffractive coupling rate, $U$.

\subsection{Gap-confined polariton states: $W(x)\neq 0$, $P(x)=0$}\label{sec:gap_states}
In this section we show numerical results obtained when an external potential is added to the model addressed in the previous section. In particular, here we focus on the energy region below the exciton resonance. Indeed, even though our model prescribes the existence of polariton bands above the exciton energy (upper polariton branches), 
we focus on experiments performed by analyzing dynamics of states below $\hbar\omega_X$, which are the lowest energy excitations and the ones towards which relaxation naturally occurs.\\
The external potential we used is given by a standard Gaussian function centered at $x=0$ 
\begin{equation}\label{eq:repulsive_potential}
V(x)=V_0 \exp\{-x^2/(2\sigma^2)\} \, ,
\end{equation}
where $V_0\geq 0$ and $\sigma$ denote the height and the standard deviation of the potential. The presence of such type of repulsive barrier can be interpreted as an extra term accounting for local changes in the refractive index for the photonic component, as well as the effects of a local blueshift of the excitonic states. In experiments, both such effects are related to the presence of the external input source $P(x)$. Here, we consider the effects of such a barrier in order to provide a clearer interpretation of the results shown in the next section, with the aim of  characterizing the structure and physical properties of eigenmodes.

Due to the local nature of $V(x)$, states with different $k$ get mixed and the band structure gets modified. In particular, our analysis shows that discrete levels appear within the energy gap between $\lambda_{+,-}(0)$ and $\lambda_{-,-}(0)$. The properties of such states have been determined by solving numerically the following eigenvalue problem:
\begin{equation}\label{eq:eigen_problem}
    \tilde{H}(x)\vec{\psi}_{n}(x)=-i\frac{\,\lambda_{n}}{\hbar}\vec{\psi}_{n}(x),\quad \lambda_n\equiv E_n-i \hbar \gamma_n
\end{equation}
where $E_n\equiv Re[\lambda_n]$ and $\hbar \gamma_n\equiv -Im[\lambda_n]$ denote the real and imaginary parts of energy eigenvalues corresponding to the eigenvectors $\vec{\psi}_{n}(x)$, respectively. 

\begin{figure}[t]
    \centering
    \includegraphics[scale=0.5]{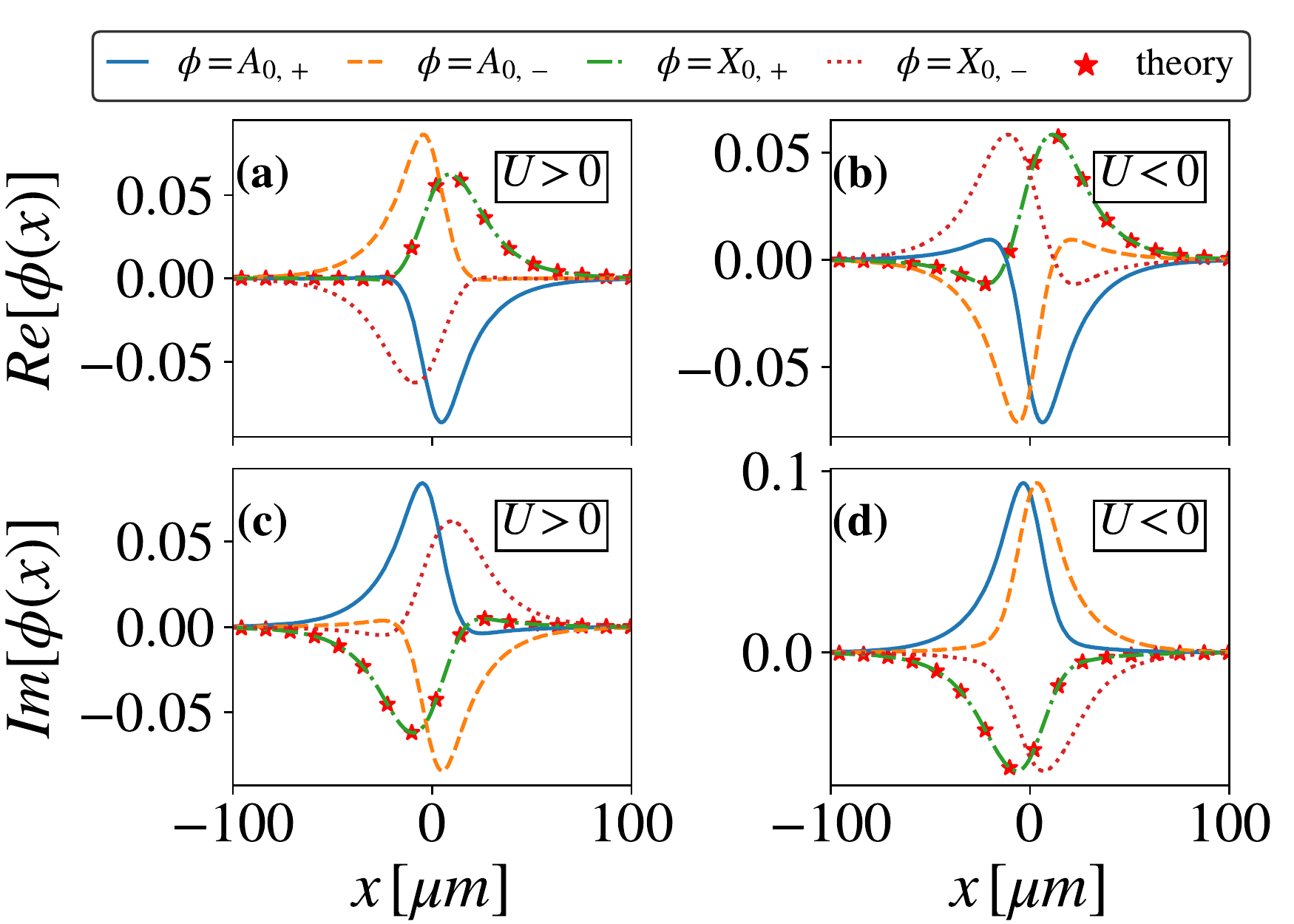}
    \caption{Spatial behavior of real ($Re[\phi]$) and imaginary ($Im[\phi]$) parts of the 4 components of $\vec{\psi}_0(x)$, for a Gaussian potential with $V_0=5$ meV and $\sigma=10$ $\mu$m. Panels (a) and (c) describe data for $\hbar U=4.45$ meV, while panels (b) and (d) correspond to $\hbar U=-4.45$ meV. The star points labeled as ``theory'' are a plot of Eq.~\eqref{eq:excitonic_vs_photonic_components} obtained from the numerical solution for $A_{0,\,+}(x)$ (a similar agreement is observed for the other component, not shown). Numerical results are obtained for $\hbar \gamma_A=\hbar \tilde{\gamma}_A=\hbar \gamma_X=0.1$ meV. The other relevant parameters are set as in Fig.~\ref{fig:pol_bands_example}. }
    \label{fig:example_eigenfunction0}
\end{figure}

\begin{figure}[t]
    \centering
    \includegraphics[scale=0.5]{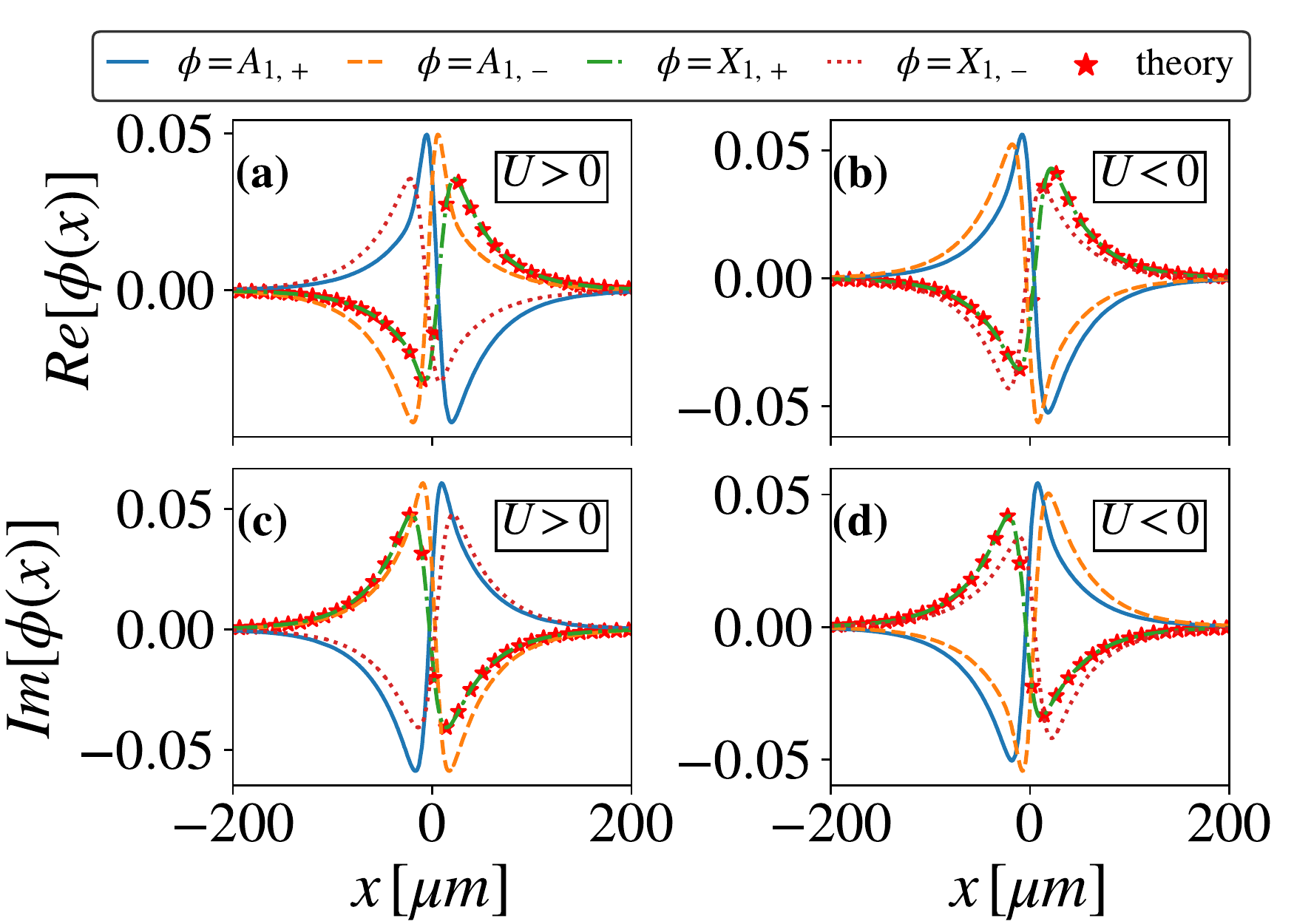}
    \caption{Spatial behavior of real ($Re[\phi]$) and imaginary ($Im[\phi]$) parts of the 4 components of $\vec{\psi}_1(x)$, calculated for the very same parameters assumed for the results of Fig.~\ref{fig:example_eigenfunction0}.
    Again, panels (a) and (c) describe data for $\hbar U=4.45$ meV, while panels (b) and (d) correspond to $\hbar U=-4.45$ meV. The star points labeled as ``theory'' are a plot of Eq.~\eqref{eq:excitonic_vs_photonic_components} obtained from the numerical solution for $A_{1,\,+}(x)$, with a similar agreement obtained for the other component (not shown). }
    \label{fig:example_eigenfunction1}
\end{figure}

As an example, we report in Figs.~\ref{fig:example_eigenfunction0} and \ref{fig:example_eigenfunction1} the numerical results of Eq.~\eqref{eq:eigen_problem} for the first two eigenmode components of $\vec{\psi}_{n}(x)$ (i.e., for $n=0$ and $n=1$, respectively). 
As a benchmark for the numerical solution, we notice that 
\begin{equation}\label{eq:excitonic_vs_photonic_components}
X_{n,\,\pm}(x)=\frac{-i \Omega_R\,A_{n,\,\pm}(x)}{-i\frac{\lambda_{n}}{\hbar}+i\left(\omega_X+i \gamma_X+ \frac{V(x)}{\hbar}\right)} \, ,
\end{equation}
as it is derived by direct inspection of the eigenvalue problem in Eq.~\eqref{eq:eigen_problem}.
Hence, we also plot in Figs.~\ref{fig:example_eigenfunction0} and \ref{fig:example_eigenfunction1} the behavior of the $X_{n,\,+}(x)$ component obtained by plugging the numerical solution corresponding to $A_{n,\,+}(x)$ into Eq.~\eqref{eq:excitonic_vs_photonic_components} (markers labeled as ``theory"), which perfectly match the numerically computed solutions for the same component.  

In addition, our numerical analysis also suggests that the photonic components $A_{n,\,+}(x)$ and $A_{n,\,-}(x)$ are connected by the inversion operation, that is $x \to -x$ (see App.~\ref{app:symm_gap_states} for details). In particular, a different symmetry is obtained for positive and negative values of $U$. For $U>0$ data obtained for $\overrightarrow{\psi}_{n}$ suggest that the photonic components are related by the following expressions
\begin{equation}\label{eq:spatial_symmetry_Bic}
    A_{n,\,+}(x)=(-1)^{n+1}A_{n,\,-}(-x),
\end{equation}
while for $U<0$ the relation becomes 
\begin{equation}\label{eq:spatial_symmetry_Lossy}
    A_{n,\,+}(x)=(-1)^{n}A_{n,\,-}(-x) \, .
\end{equation}
Interestingly, given the expression reported in Eqs.~\eqref{eq:excitonic_vs_photonic_components}, \eqref{eq:spatial_symmetry_Bic} and \eqref{eq:spatial_symmetry_Lossy}, it is easy to verify that 
\begin{equation}\label{eq:orthogonality_even_odd}
\int dx 
\langle 
\overrightarrow{\psi}_{n} (x),\overrightarrow{\psi}_{n+(2l+1)}(x)\rangle = 0
\end{equation}
for $n$ and $l$ integers, i.e., 
any two eigenstates  $\overrightarrow{\psi}_{n}$ and $\overrightarrow{\psi}_{m}$ are orthogonal whenever $n$ and $m$ have an opposite parity.

\begin{figure}[t]
\centering
\includegraphics[scale=0.5]{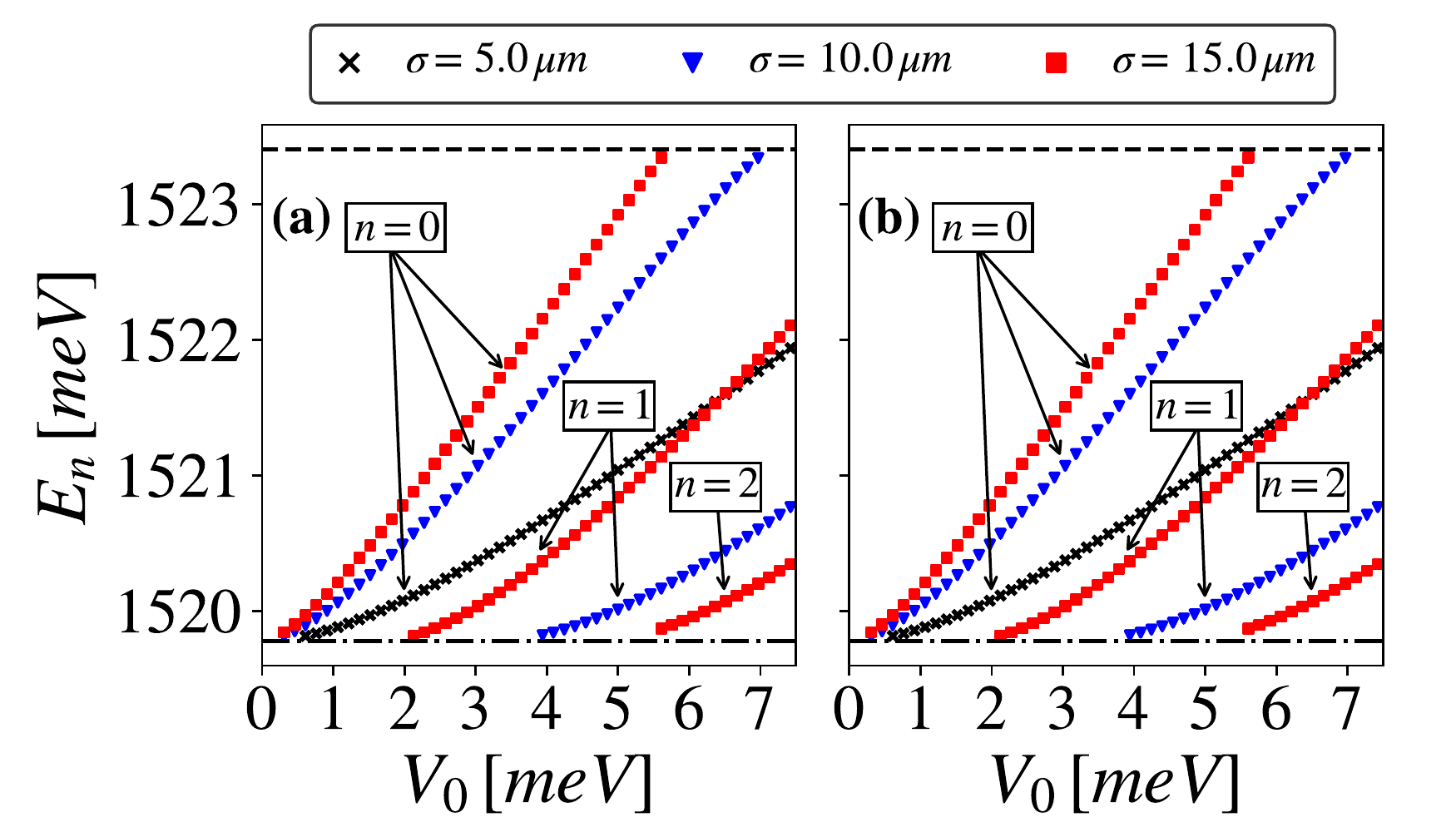}
\caption{Real part of the discrete levels eigenenergies, $E_n$ (in meV) created by the repulsive potential within the polariton gap below the exciton resonance, as a function of the potential height $V_0$ and for $\sigma=5,\,10,\,15$ $\mu$m (see legend), respectively. Results are shown for (a) $U>0$ and (b) $U<0$, respectively. The relevant model parameters are set as in Fig.~\ref{fig:pol_bands_example}. In both panels, the dot-dashed horizontal line below 1520 meV, and the dashed horizontal line above 1523 meV correspond to the band extrema of the eigenmodes $\lambda_{+,\,-}(k)$ (maximum at $k=0$) and  $\lambda_{-,\,-}(k)$ (minimum), respectively (see also Fig.~\ref{fig:pol_bands_example}(a)). }\label{fig:real_eigenvalues}
\end{figure}

We proceed by showing the behavior of the eigenenergies corresponding to $\vec{\psi}_{n}(x)$ within the lower polariton gap as a function of the parameter $V_0$ and for different values of $\sigma$. The numerical results are shown in Fig.~\ref{fig:real_eigenvalues}. Numerical data reported in Fig.~\ref{fig:real_eigenvalues}(a) describe the behavior for $\hbar U=4.45$ meV, while in Fig.~\ref{fig:real_eigenvalues}(b) are reported data corresponing to $\hbar U=-4.45$ meV.\\
Similarly to what observed for $Real[\lambda]$, the energy of such discrete levels does not depend on the sign of $U$. So for the sake of simplicity, let us pay attention to Fig.~\ref{fig:real_eigenvalues}(a). In analogy to the behavior expected when considering a confining potential $\tilde{V}(x)=-V(x)$ perturbing the motion of a positive-effective mass  ($m^*$) quantum particle, that is 
\begin{equation}\label{eq:schrodinger}
-\frac{\hbar^2}{2m^*}\frac{d^2}{dx^2}\phi_n(x) + \tilde{V}(x)\phi_n(x) = E_n\phi_n(x),\quad m^*>0
\end{equation}
the number of bound-states as well as the distance with respect to the minimum of the parabolic band $E(k)={\hbar^2k^2}/{(2m^*)}$, only depend on the properties of the potential well $\tilde{V}(x)$, namely its width and depth. Obviously, the wider the well, the larger the number of bound states supported by the well. A similar behavior is observed in the present case, where negative effective mass excitations are trapped within a  potential barrier. For $\sigma=5$ $\mu$m, the potential supports a single discrete state (i.e., with $n=0$). For $\sigma=10$ $\mu$m, in the range of $V_0$ values considered, the system supports two discrete levels, the second entering in the gap in the vicinity of $V_0\approx 4$ meV. When the width of the repulsive barrier is increased further, also the number of discrete states increases. This behavior is confirmed by the results for $\sigma=15$ $\mu$m, where the second and third discrete levels supported by the repulsive potential appear in the vicinity of $V_0\approx 2$ meV and $V_0\approx 5.5$ meV.\\
The first main difference with respect to the solutions of Eq.~\eqref{eq:schrodinger} concerns the range of energies spanned by such bound states while sweeping $V_0$. In the standard positive mass case, the deeper the potential, the smaller the quantization energy. In the present case, as it is obtained in Fig.~\ref{fig:real_eigenvalues}, the eigenenergies of such states are always bounded between a lower and an upper value. Such limits correspond to the maximum of the negative-mass band $\lambda_{+,\,-}$ and to the minimum of the positive-mass band $\lambda_{-,\,-}$, respectively. In particular, when a state reaches and crosses the minimum of the upper band at $\lambda_{-,\,-}$, it can no longer be normalized and it represents an extended solution across the whole real space domain.

\begin{figure}[t]
\centering
\includegraphics[scale=0.5]{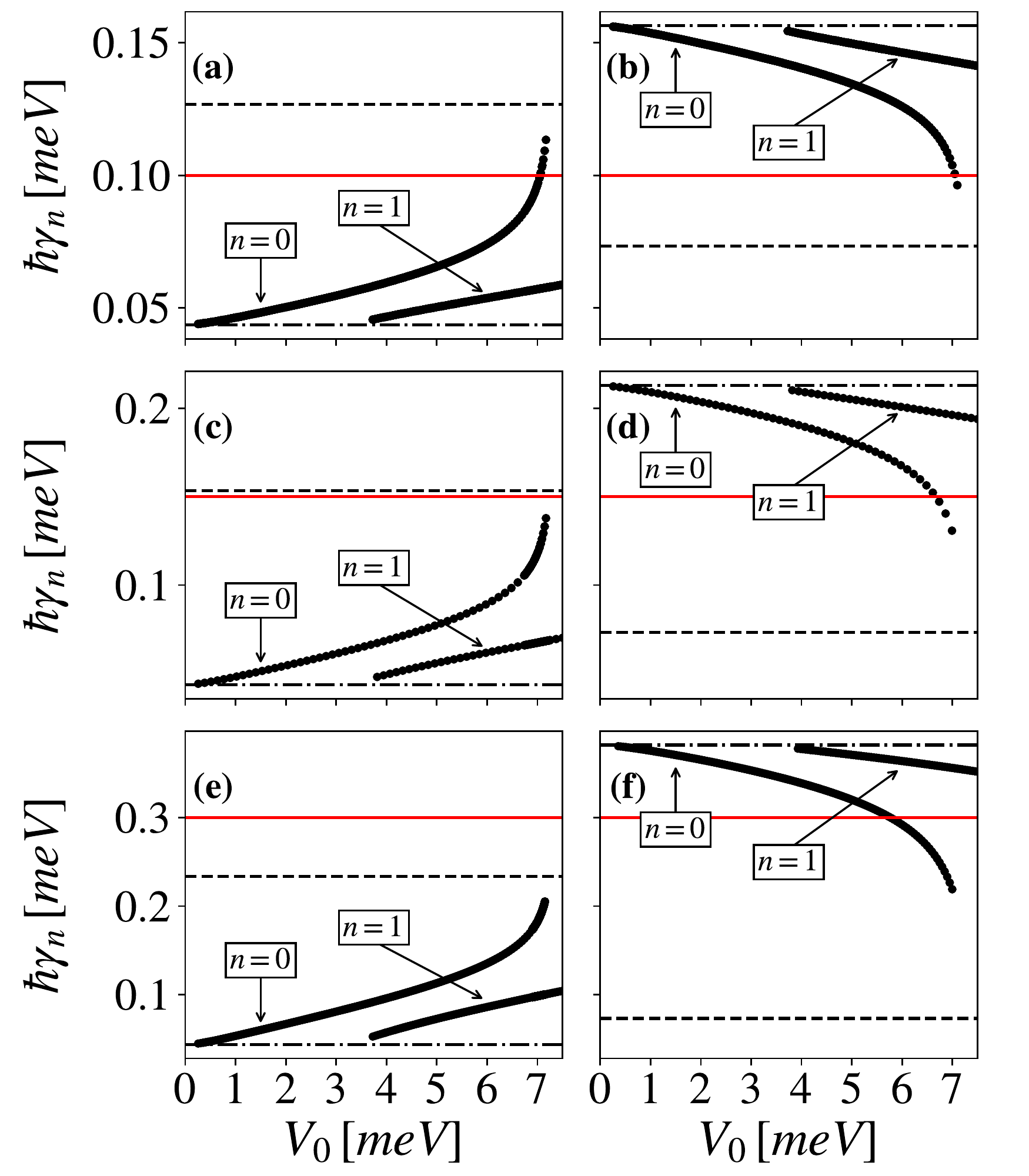}
\caption{Imaginary parts of discrete eigenmodes within the energy gap, $\hbar \gamma_n$ (in meV) plotted as a function of $V_0$ for $\sigma=10$ $\mu$m. Results for $\hbar U = 4.45$ meV are shown for (a) $\hbar \gamma_A=\hbar\tilde{\gamma}_A=0.1$ meV, (c) $0.15$ meV, and (e) $0.3$ meV, respectively.
Results for $\hbar U = -4.45$ meV are shown for (b) $\hbar \gamma_A=\hbar\tilde{\gamma}_A=0.1$ meV, (d) $0.15$ meV, and (f) $0.3$ meV, respectively. In these simulations we have assumed $\hbar \gamma_X=0.1$ meV for the exciton modes. All the other parameters are set as in Fig.~\ref{fig:pol_bands_example}. In all the panels: the dot-dashed horizontal line corresponds to the imaginary part of the $\lambda_{+,-}(k=0)$ eigenmode, the dashed horizontal line corresponds to the one of $\lambda_{-,-}(k=0)$, the solid horizontal line is placed at the value $\hbar \gamma_A$.}\label{fig:imag_eigenvalues}
\end{figure}

While the resonant energy (real part of the eigenvalues) within the energy gap is not affected by the sign of $U$, the losses (i.e., the imaginary part of the eigenvalues) of such discrete levels strongly depend on the nature of the negative mass branch giving out of which they arise on increasing $V_0$. In fact, numerical results for either $U>0$ or $U<0$ are reported in Fig.~\ref{fig:imag_eigenvalues}, and for different values of $\hbar\gamma_A$. In particular, results are seen to be quite different depending on the sign of $U$, displaying low qualitative dependence on the value of $\gamma_A$. We thus focus on commenting the differences between Figs.~\ref{fig:imag_eigenvalues}(a) and \ref{fig:imag_eigenvalues}(b). When $V_0$ is slightly larger than zero, the first mode entering in the gap ($n=0$) is characterized by an imaginary part $\hbar\gamma_n$ that is close to the one corresponding to the polariton band $\lambda_{+,-}(k=0)$ (dot-dashed horizontal line in both panels). On the increasing of $V_0$, deviations from this value are observed. The latter depends on the sign of $U$ (see also Fig.~2). Hence, when $U>0$ an increasing  of the repulsive barrier height yields an increased $\gamma_n$, as seen in Fig.~\ref{fig:imag_eigenvalues}(a).  Conversely, when the system is characterized by a negative diffractive coupling $U$, $\gamma_n$ decreases on increasing $V_0$, as shown in Fig.~\ref{fig:imag_eigenvalues}(b).  In particular, similarly to what already shown for the real part of the eigenvalues, also the imaginary parts of the discrete gap-confined levels are bound to vary between those of the negative- and positive-mass bands at $k=0$, i.e.,  $\lambda_{+,-}(k=0)$ and $\lambda_{-,-}(k=0)$, respectively.\\
In summary: the discrete levels created by the repulsive potential $V(x)$ are confined within the polariton energy-gap, and their complex eigenvalues are such that real and imaginary parts vary continuously between those of $\lambda_{+,-}(k=0)$ and $\lambda_{-,-}(k=0)$.

\section{Two-mode case: polariton condensation in the CW regime}\label{sec:pol_cond_two}
As it is known from previous studies, polariton condensation is identified from accumulation of excitations in one specific state, usually the lowest energy one in planar microcavities, and this phenomenology is well captured by a one-dimensional single-mode GPE formulation. In this Section we show  numerical results concerning the onset of polariton condensation in the steady-state regime, when the coupled multi-band exciton-photon system is connected to a particle reservoir driven by a spatially-dependent CW input source. The expression of such driving term is given as
\begin{equation}\label{eq:pump}
P(x)=P_0 \exp\{-x^2/(2\sigma^2)\} \, ,
\end{equation}
in which $P_0$ is a time-independent pump-rate per $\mu$m, i.e., $[P_0]=$ ($\mu$m$\cdot$ps)$^{-1}$. In particular, following the discussion from the previous Section, it is hereby assumed that the pumping rate is linearly related to the barrier height, $V_0$ by the following relation:
\begin{equation}\label{eq:pump_vs_conf}
P_0=\frac{\eta V_0}{\hbar} \, ,
\end{equation}
in which $\eta$ is a parameter having the dimension $[\eta]=\mu$m$^{-1}$. Concerning the reservoir-polariton interactions, we assume the two excitonic modes to be similarly coupled to the reservoir density, and we made the same assumption for the photonic components for simplicity of computation and without loss of generality. In this scenario,  the operator $G$ is diagonal and, due to the constraints reported in Eqs.~\eqref{eq:G-constraint-positivity} and \eqref{eq:G-constraint-normalization}, it has the following structure:
\begin{equation}\label{eq:G_mat_simulations}
G=\left(
\begin{array}{cccc}
(1-\alpha)/2 & 0 & 0  & 0\\
0 & (1-\alpha)/2 & 0 & 0\\
0 & 0 & \alpha/2 & 0\\
0 & 0 & 0 & \alpha/2\\
\end{array}\right)
\end{equation}
with $\alpha$ being an effective parameter such that $0 \leq \alpha\leq 1$.

Since we are interested in the stationary solutions of Eq.~\eqref{eq:dynamical_system} at large time $t$, and since the model always admits as a possible solution $\vec{\psi}(x)=0$, we consider an initial configuration characterized by a small population. By doing so, we are able to understand whether a tiny perturbation added to the ``empty'' solution  actually leads to the appearance of a stable, macroscopically populated condensate. Details about the system initialization are given in App.~\ref{app:sys_init}.
Such initial configuration is then evolved in time by means of a standard numerical integration algorithm (i.e., explicit Runge-Kutta 4(5) method), and the convergence towards a possibly nonempty condensate configuration is monitored by looking at the time behavior of the total population, that is
\begin{equation}\label{eq:density_light_matter}
\begin{split}
&N_{\psi}(t)= \int dx \langle \vec{\psi}(x,\,t),\vec{\psi}(x,\,t)\rangle=\\
&=\int dx \left[\,\vert A_{+} \vert^2 +\vert A_{-} \vert^2+\vert X_{+} \vert^2+\vert X_{-} \vert^2\right],
\end{split}
\end{equation}
where $A_{\sigma}\equiv A_{\sigma}(x,\,t)$ and $X_{\sigma}\equiv X_{\sigma}(x,\,t)$.\\

\begin{figure}[t]
\centering 
\includegraphics[scale=0.5]{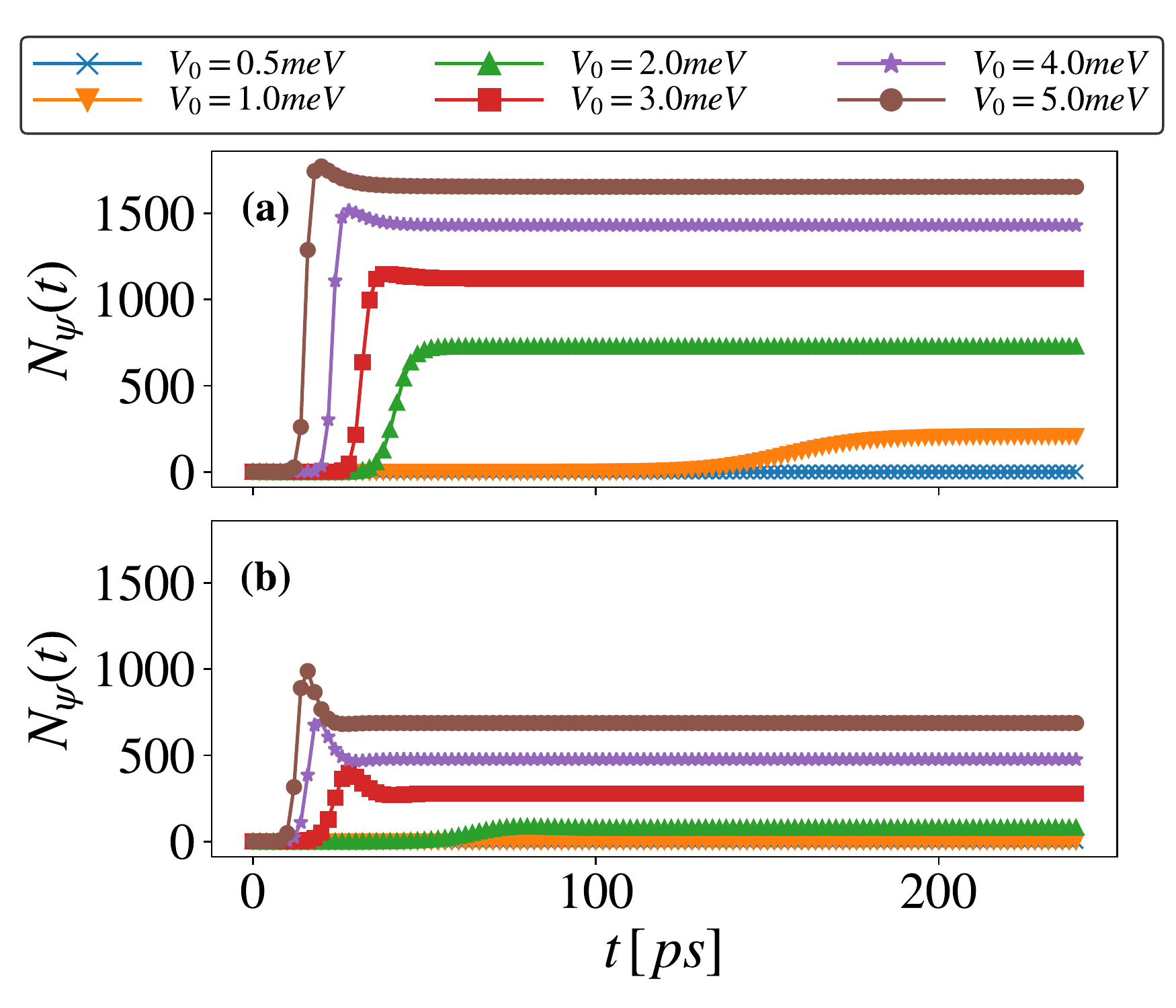}
\caption{Behavior of the condensate occupation $N_{\psi}(t)$ (Eq.~\eqref{eq:density_light_matter}) as a function of time (in ps) for different values of $V_0$ (top legend), for (a) $\hbar U=4.45$ meV and (b) $\hbar U=-4.45$ meV. Results are reported for $\hbar \gamma_A=\hbar \tilde{\gamma}_A=0.1$ meV, $\hbar \gamma_X=0.1$ meV, $\alpha=0.01$ (Eq.~\eqref{eq:G_mat_simulations}), $\eta=2$ $\mu$m$^{-1}$ and $g=0.1$ $\mu$m/ps. All the other relevant parameters are set as in Fig.~\ref{fig:pol_bands_example}.}\label{fig:convergence_to_ss}
\end{figure} 

An example of time evolution is reported in Fig.~\ref{fig:convergence_to_ss}, where the behavior of $N_{\psi}(t)$ as a function of time is again compared for the cases $U>0$ and $U<0$, assuming the same CW driving protocol and the same values for $V_0$. Even if the initial population injected in the light-matter subsystem is small (see App.~\ref{app:sys_init}), independently of the sign of $U$, for sufficiently large values of $V_0$ the system dynamics tends towards a stable steady-state configuration displaying a macroscopic occupation, i.e.,
\begin{equation}
\lim_{t\to\infty} N_{\psi}(t)= N_{\psi,\,ss} \gg 0.
\end{equation}

Interestingly, according to the results shown in Fig.~\ref{fig:convergence_to_ss}, the convergence towards a nonempty condensate seems to depend on the sign of $U$. Indeed, if on the one hand for $V_0>1$ meV in both cases the system approaches a configuration with $N_{\psi,\,ss} >0 $, on the other hand at exactly $V_0=1$ meV polariton condensation is only triggered for $U>0$. \\
In particular, since $V_0$ is in one-to-one correspondence with $P_0$ (Eq.~\eqref{eq:pump_vs_conf}), such behavior suggests that in systems with $U>0$ condensation would occur for lower values of $P_0$, i.e., they would display a lower condensation threshold. An interesting prediction to be verified experimentally.

\begin{figure}[t]
\centering
\includegraphics[scale=0.5]{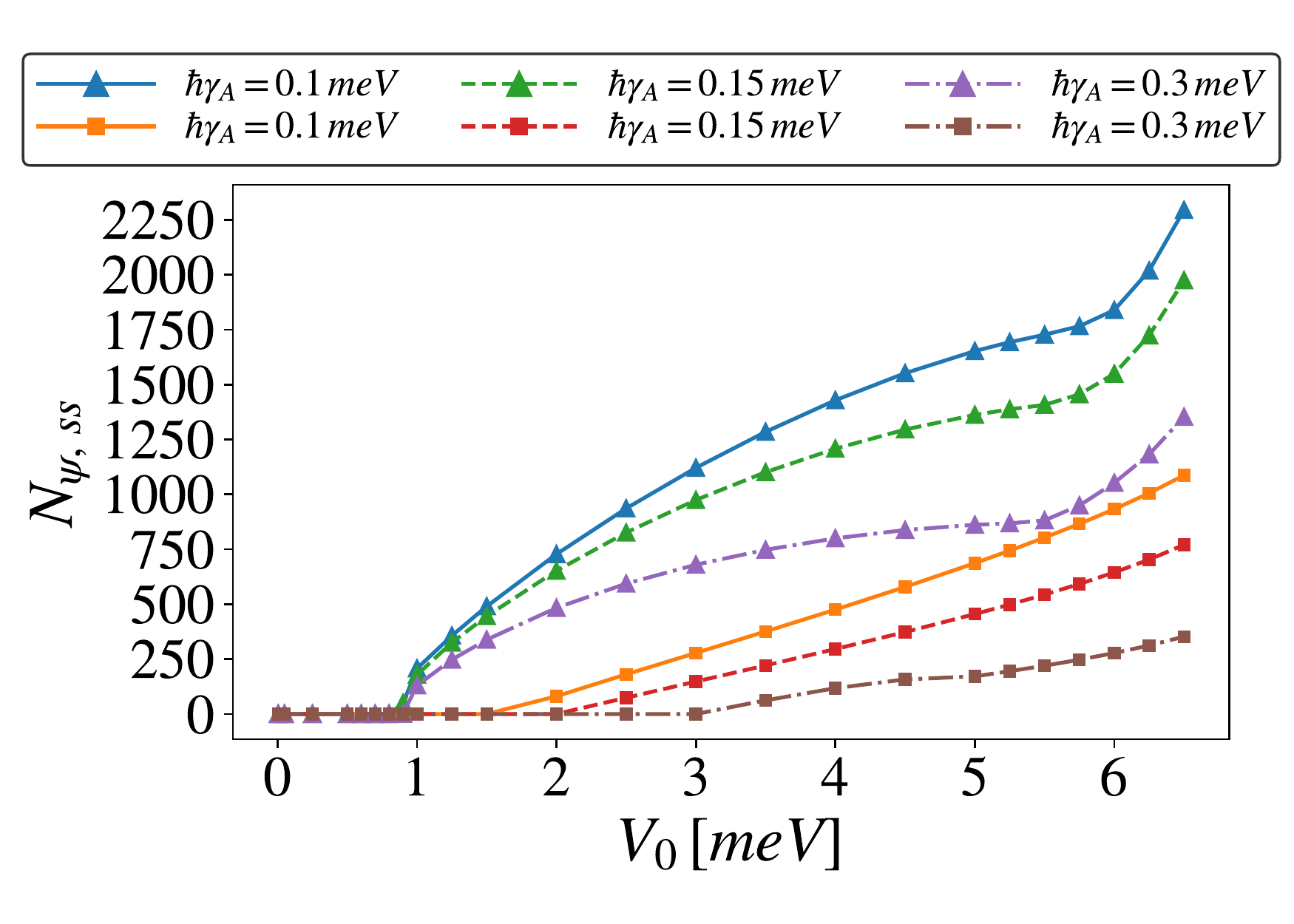}
\caption{Steady-state occupation of the exciton-photon coupled system, $N_{\psi\,ss}$, as a function of $V_0$ (in meV), for different values of $\hbar \gamma_A$ (top legend). Triangle markers correspond to data obtained for $\hbar U=4.45$ meV, while square points are for $\hbar U = -4.45$ meV. All the other relevant parameters are set as in Fig.~\ref{fig:convergence_to_ss}.}\label{fig:ss_occupation}
\end{figure}

In order to understand whether this is indeed the case, we performed a characterization of the dependence of the steady-state occupation $N_{\psi,\,ss}$ on $V_0$. Results obtained for different values of the photonic losses are shown in Fig.~\ref{fig:ss_occupation}. As in the previous cases, we compare the trend in data obtained for $U>0$ (triangles) and $U<0$ (squares), respectively. For what concerns the condensation threshold, data corresponding to $U>0$ do not display a significant dependence on $\hbar \gamma_A$. In particular, for all the cases considered the critical value of $V_0$ leading to condensation is slightly below 1 meV. On the contrary, at first glance, numerical results for $U<0$ display a considerably different behavior. In particular, the condensation threshold in this case shows more pronounced dependence on the photonic losses, going from $V_0 \simeq 1$ meV to $\simeq 3$ meV while increasing $\hbar\gamma_A$ from $0.1$ to $0.3$ meV.

In addition, data obtained for values of $U$ having a different sign show a quantitatively different dependence on $V_0$. If on the one hand they clearly show that condensation is occurring in both cases, on the other hand the value of $U$ does not only affect the condensation threshold.
In the next two sections we show that all these peculiar features of the model can be interpreted by looking at the properties of the states previously characterized in Sec.~\ref{sec:gap_states}. In particular, we first address the behavior of the losses of the condensed states. Then, we analyse the spectral properties of the radiation emitted from the condensate, and we show that it is peaked at energies corresponding to the discrete levels considered in the previous section.

\subsection{Analysis of the condensate losses}\label{sec:anal_linewidth}
In this section, by analyzing the condensate losses, we show that the steady-state configuration is given by one or few of the discrete eigenmodes characterized in the previous section. However, in order to keep the discussion as clear as possible and to show the main idea, we first assume that the configuration approached by the system during relaxation towards its lower energy eigenstates corresponds to a single normalized eigenfunction $\vec{\psi}_n(x)$ of a gap-confined state, as obtained from the operator $\tilde{H}(x)$ characterized in Sec.~\ref{sec:gap_states}. In this case, the steady state of the system is given by 
\begin{equation}
    \lim_{t\to +\infty} \vec{\psi}(x,\,t)=\sqrt{N_{\psi,ss}}\vec{\psi}_{n}(x) \, .
\end{equation}
Then, let us consider the spatial integral of Eq.~\ref{eq:total_density_evolution}, with the steady-state condition imposing that the derivative of the total particle density should become zero. Eq.~\ref{eq:total_density_evolution} then reduces to  
\begin{equation}\label{eq:reservoir_ss_occupation}
\begin{split}
0&=\int P(x)\,dx-\frac{1}{\tau_R}\int  n_{ss}(x)\,dx-2 \gamma_{n} N_{\psi,\,ss}=\\
&=\int P(x)\,dx-\frac{1}{\tau_R}N_{R,\,ss}-2 \gamma_{n} N_{\psi,\,ss} \, ,
\end{split}
\end{equation}
in which $N_{R,\,ss}$ denotes the total number of particles accumulated in the reservoir. Since $N_{\psi,\,ss}>0$, we can rearrange the terms in Eq.~\eqref{eq:reservoir_ss_occupation} and get an expression for the imaginary part of the n-th mode steady state, $\gamma_{n}$, which is 
\begin{equation}\label{eq:linewidth_ss}
\gamma_{n}=\frac{\int P(x)\,dx-\gamma_R\,N_{R,\,ss}}{2\,N_{\psi,\,ss}} \, ,
\end{equation} 
in which $\gamma_R=1/\tau_R$. By plugging the expressions reported in Eqs.~\eqref{eq:pump} and \eqref{eq:pump_vs_conf} into Eq.~\eqref{eq:linewidth_ss}, we finally obtain an analytic expression for the condensate losses as
\begin{equation}\label{eq:theoretical_linewidth}
\hbar \gamma_{n}=\frac{\eta V_0 \sqrt{2\pi}\sigma-\hbar\gamma_R\,N_{R,\,ss}}{2\,N_{\psi,\,ss}} \, .
\end{equation}
In general, depending on the values of $V_0$ and $\sigma$ in the expression for the repulsive potential, the stationary configuration might correspond to a superposition of more than one gap-confined eigenstate. For instance, as shown in Fig.~\ref{fig:real_eigenvalues}, when $\sigma=10$ $\mu$m and $V_0\in[0,\,7]$ meV, two discrete levels appear within the polariton gap, whose eigenfunctions are $\psi_0(x)$ and $\psi_1(x)$. Then, let us suppose that 
\begin{equation}
    \vec{\psi}_{ss}(x)=\alpha_0 \vec{\psi}_{0}(x)+ \alpha_1\vec{\psi}_{1}(x) \,  .
\end{equation}
Since the two eigenstates are orthogonal by symmetry constraints, see Eq.~\eqref{eq:orthogonality_even_odd}, by following the same steps leading to Eq.~\eqref{eq:theoretical_linewidth} we can write the overall steady state condensate loss in terms of the weighted imaginary parts of the two discrete modes that are present in the gap as
\begin{equation}\label{eq:losses_superposition}
    \hbar \bar{\gamma}=\frac{\hbar\gamma_0 \vert\alpha_0\vert^2+\hbar\gamma_1 \vert\alpha_1\vert^2}{N_{\psi,\,ss}}=\frac{\eta V_0 \sqrt{2\pi}\sigma-\hbar\gamma_R\,N_{R,\,ss}}{2\,N_{\psi,\,ss}} \, ,
\end{equation}
in which $N_{\psi,\,ss}= \vert \alpha_0\vert^2+\vert \alpha_1\vert^2$.

\begin{figure}[t]
\centering
\includegraphics[scale=0.5]{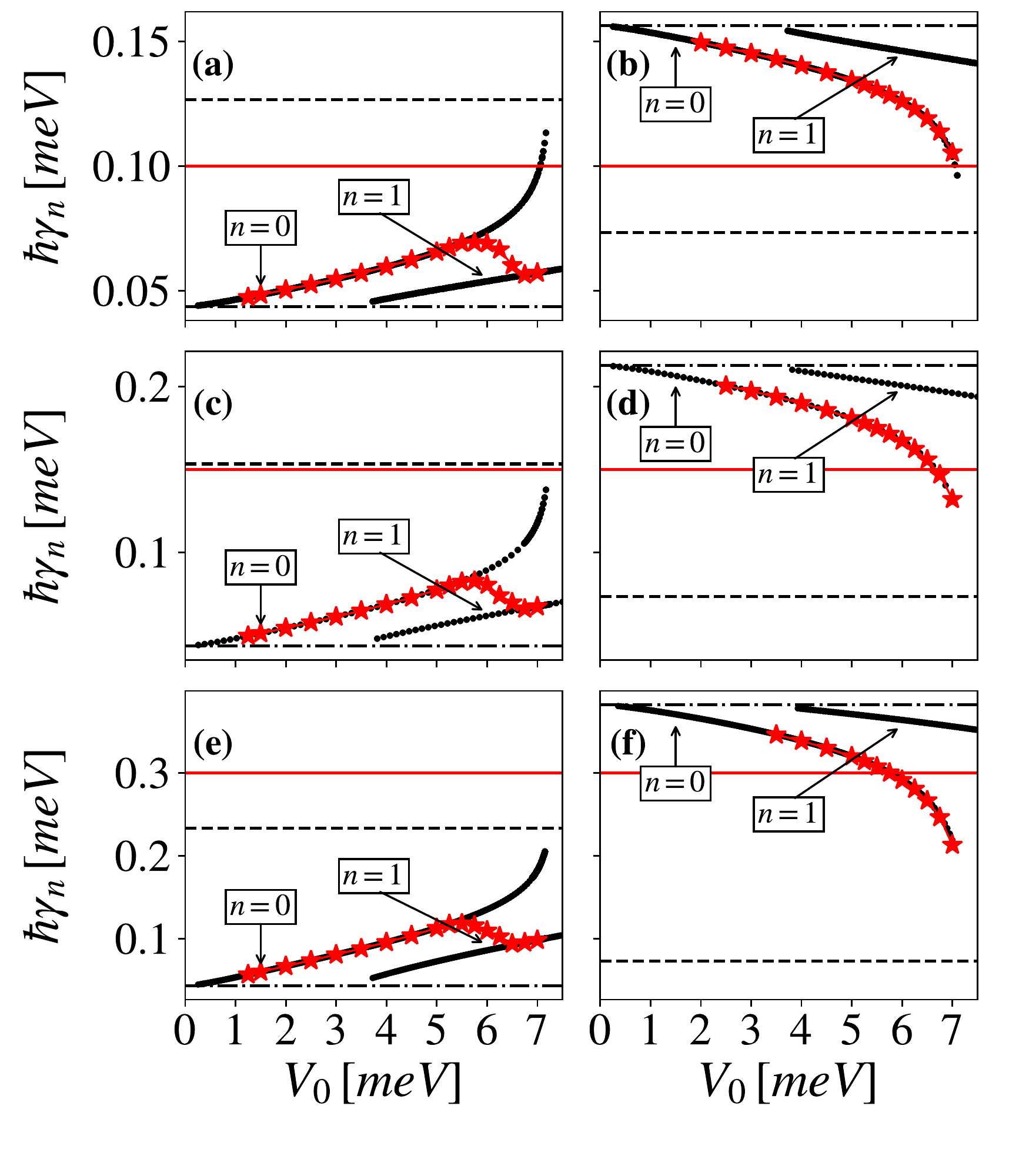}
\caption{Comparison between the imaginary parts of the discrete eigenmodes induced by the potential $V(x)$ ($\hbar\gamma_n$) and the expected condensate losses obtained from Eq.~\eqref{eq:losses_superposition}, 
(star-shaped markers), plotted as a function of $V_0$. Results are shown for $U>0$ and (a) $\hbar \gamma_A=\hbar \tilde{\gamma}_A=0.1$ meV, (c) 0.15 meV, and (e) 0.3 meV, respectively (solid horizontal line). Similarly, results are shown for $U<0$ and (b) $\hbar\gamma_A=\hbar \tilde{\gamma}_A=0.1$ meV, (d) 0.15 meV, and (f) 0.3 meV, respectively (solid horizontal line). The other parameters are set as in Fig.~\ref{fig:imag_eigenvalues}. The dashed and dot-dashed horizontal lines correspond to the imaginary parts of the two gap-delimiting eigenmodes, 
$\lambda_{-,-} (k=0)$ and 
$\lambda_{+,-} (k=0)$.
}\label{fig:comparison_gamma_results}
\end{figure}

{A direct comparison between Eq.~\eqref{eq:losses_superposition} and the numerical results previously reported in Fig.~\ref{fig:imag_eigenvalues} is given in Fig.~\ref{fig:comparison_gamma_results}. For $U<0$, as shown in Figs.~\ref{fig:comparison_gamma_results}(b,d,f) for different values of $\gamma_A=\tilde{\gamma}_A$, the steady-state condensate loss obtained from of Eq.~\eqref{eq:losses_superposition} (star-shaped markers) nicely follow the numerical solutions for the imaginary parts of the  first discrete level appearing within the polariton band gap on increasing $V_0$, for all the values of $V_0$ such that $N_{\psi,\,ss}> 0$ (see, e.g., Fig.~\ref{fig:ss_occupation}). In particular, in this case the imaginary part of eigenmodes decreases when increasing $V_0$, and we clearly notice that our protocol leads to the stabilization of a configuration compatible with the macroscopic occupation of the first mode confined in the gap. \\
A similar behavior is displayed for $U>0$, results shown in  Figs.~\ref{fig:comparison_gamma_results}(a,c,e).  However, by further increasing the pumping rate (i.e., $V_0$ in our model) the star-shaped markers move along a path going smoothly from the branch corresponding to the first gap-confined eigenmode imaginary part to the one corresponding to the second discrete eigenmode. Such behavior is somehow compatible with the fact that the first discrete state of the potential is moving towards the upper polariton branch, $\lambda_{-,-}(k)$, and it is getting out from the energy gap (see also Fig.~\ref{fig:real_eigenvalues}), such that also the second mode within the gap becomes populated, as it will be detailed in the following Section.}

\subsection{Radiation emission from the condensate}\label{sec:light_emission}
\begin{figure}[t]
\centering
\includegraphics[scale=0.5]{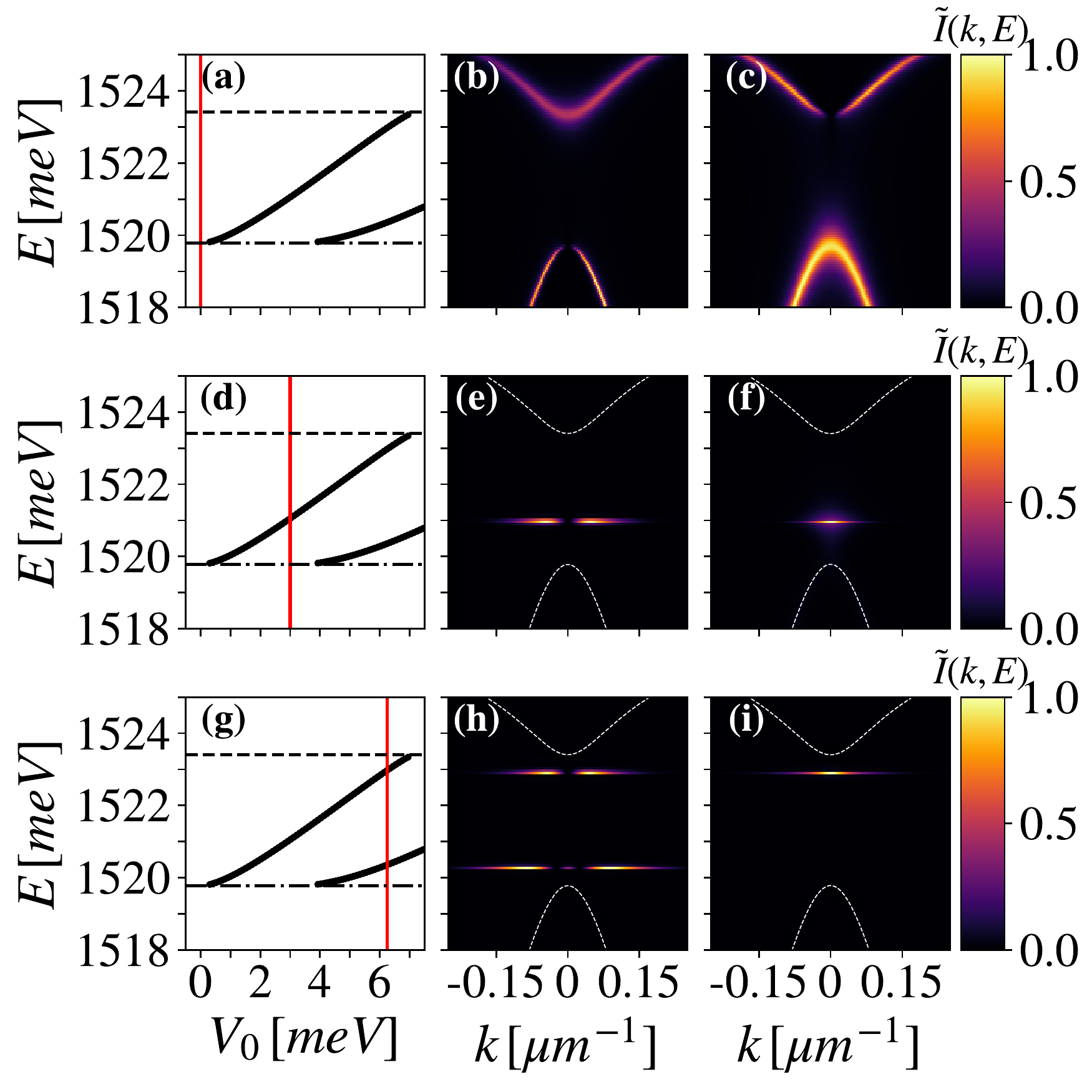}
\caption{Direct comparison between the energy of the discrete levels created by the potential $V(x)$ within the polariton gap (real part of eigenvalues, coinciding for either positive or negative $U$, as shown in panels a, d, g) calculated for $\sigma=10$ $\mu$m and $\hbar\gamma_A=\hbar\tilde{\gamma}_A=0.3$ meV. The corresponding (normalized) spectral density obtained from Eq.~\ref{fig:spectral_density} is shown for $\hbar U=4.45$ meV and corresponding to the three values of $V_0$  highlighted with the vertical red lines in panels (a,d,g), respectively, in panels (b) $V_0=0$, (e) $V_0=3.0$ meV, and (h) $V_0=6.25$ meV. Similarly, results are shown for $\hbar U=-4.45$ meV in panels (c), (f), and (i) in correspondence of the same values of $V_0$. All the other relevant parameters are set as in Fig.~\ref{fig:convergence_to_ss}.  }\label{fig:spectral_density}
\end{figure}

In this section we show numerical results describing the energy-momentum spectral density of the the emitted radiation from the polariton system. In the hypothesis that the field emitted by the structure is proportional to the field inside the system (as it is done in a standard input-output scenario), the emitted radiation has an overall space-time profile $A(x,\,t)$ given by 
\begin{equation}\label{eq:out_field}
A(x,\,t)\propto A_{+}(x,\,t)+A_{-}(x,\,t) \, . 
\end{equation}
Then the energy-momentum characteristics of the emitted radiation are essentially encoded into the Fourier transform of $A(x,\,t)$ 
\begin{equation}\label{eq:ft_field}
A(k,\,E)=\int\frac{dk}{2\pi}\int \frac{dE}{2\pi\hbar}e^{-ikx}e^{iEt/\hbar} A(x,\,t).
\end{equation}
The corresponding spectral density is given as 
\begin{equation}\label{eq:spectral_density}
I(k,\,E)=\vert A(k,\,E) \vert^2 \, .
\end{equation}
In order to display results related to the dependence of the spectral density on $V_0$, we plot in Fig.~\ref{fig:spectral_density} the normalized quantity defined as 
\begin{equation}
\tilde{I}(k,\,E)=I(k,\,E)/\mbox{max}[I(k,\,E)] \, .
\end{equation}

We report results for the spectral density corresponding to positive $U$ and different values of the pumping strength, quantified as  $V_0=0$ in Fig.~\ref{fig:spectral_density}(b), $V_0=3.0$ meV in Fig.~\ref{fig:spectral_density}(e), and  $V_0=6.25$ meV in Fig.~\ref{fig:spectral_density}(h). The same quantity is also reported for the corresponding values of $V_0$, but assuming $U<0$ in Figs~\ref{fig:spectral_density}(c,f,i). From these results, the close connection between the discrete gap-confined eigenstates characterized in the previous Sections to the position of the peaks of the spectral density should be quite evident.

In all the cases considered, the system was initially prepared by injecting a small population within in the field $\vec{\psi}$, below the exciton states (see Eq.~\eqref{eq:initial_configuration}). Then, for each value of the parameter $V_0$ and for the two values $\hbar U = \pm 4.45$ meV, time evolution is solved until $t=100$ ps. Finally, the spectral density distribution is obtained by taking the Fourier transform along the entire space-time evolution by following the prescription reported in Eqs.~\eqref{eq:out_field}, \eqref{eq:ft_field}, and \eqref{eq:spectral_density}.\\
For $P_0=V_0=0$, the system evolves under the action of $\tilde{H}_0(x)$ (Eq.~\eqref{eq:model_no_potential}). Since no driving source as well as repulsive barrier are included into the dynamics, the spectral density displays a behavior compatible with the band structure characterized in Sec.~\ref{sec:polariton_bands} for both $U>0$ [Fig.~\ref{fig:spectral_density}(b)] and $U<0$ [Fig.~\ref{fig:spectral_density}(c)].  In particular, we notice the absence of emission at $k\simeq 0$ in the lower polariton branch evidenced in Fig.~\ref{fig:spectral_density}(b) as well as in the upper polariton branch in Fig.~\ref{fig:spectral_density}(c), which marks the behavior of a dark mode and it is compatible with recent experimental results  for the emission below threshold \cite{Ardizzone_2022, Riminucci22, Gianfrate2023}.

For $V_0=3$ meV, the system is expected to have a single discrete level within the gap. For the model employed so far, this gap-confined eigenmode is occurs at a resonant energy $E_{n=0}\simeq 1521$ meV, both for $U>0$ and $U<0$, as shown in Figs.~\ref{fig:spectral_density}(d,e,f). In fact, in this case the spectral density is essentially nonzero only in the neighborhood of this eigenmode energy, and the spectral density distribution displays a behavior as a function of the wave vector that is, again, fully compatible with experimental results \cite{Ardizzone_2022,Riminucci22,Gianfrate2023}.

Finally, by further increasing the pump-rate, a new state enters in the gap, as shown in Fig.~\ref{fig:spectral_density}(g) in which  the vertical line at $V_0=6.25$ meV crosses both branches describing the energy dependence of the first and the second gap-confined  modes supported by the structure. In agreement with the interpretation provided in the previous Section for $U>0$, when $\hbar\bar{\gamma}$ assumes an intermediate value between the first and second mode within the gap [Fig.~\ref{fig:comparison_gamma_results}(e)], the spectral density is peaked in correspondence of the energy of both states, as shown in Fig. \ref{fig:spectral_density}(h). This behavior is observe also experiments (see, e.g., Fig.2 of ``Extended data figures and table" in Ref.~\cite{Ardizzone_2022}). On the other hand, for $U<0$ we do see that the emission profile is peaked at the energy of the first discrete state within the gap, while very little population is coming from the second-order mode. This result is compatible with the data displayed in panel Fig.~\ref{fig:comparison_gamma_results}(f).

\begin{figure}[t]
\centering
\includegraphics[scale=0.45]{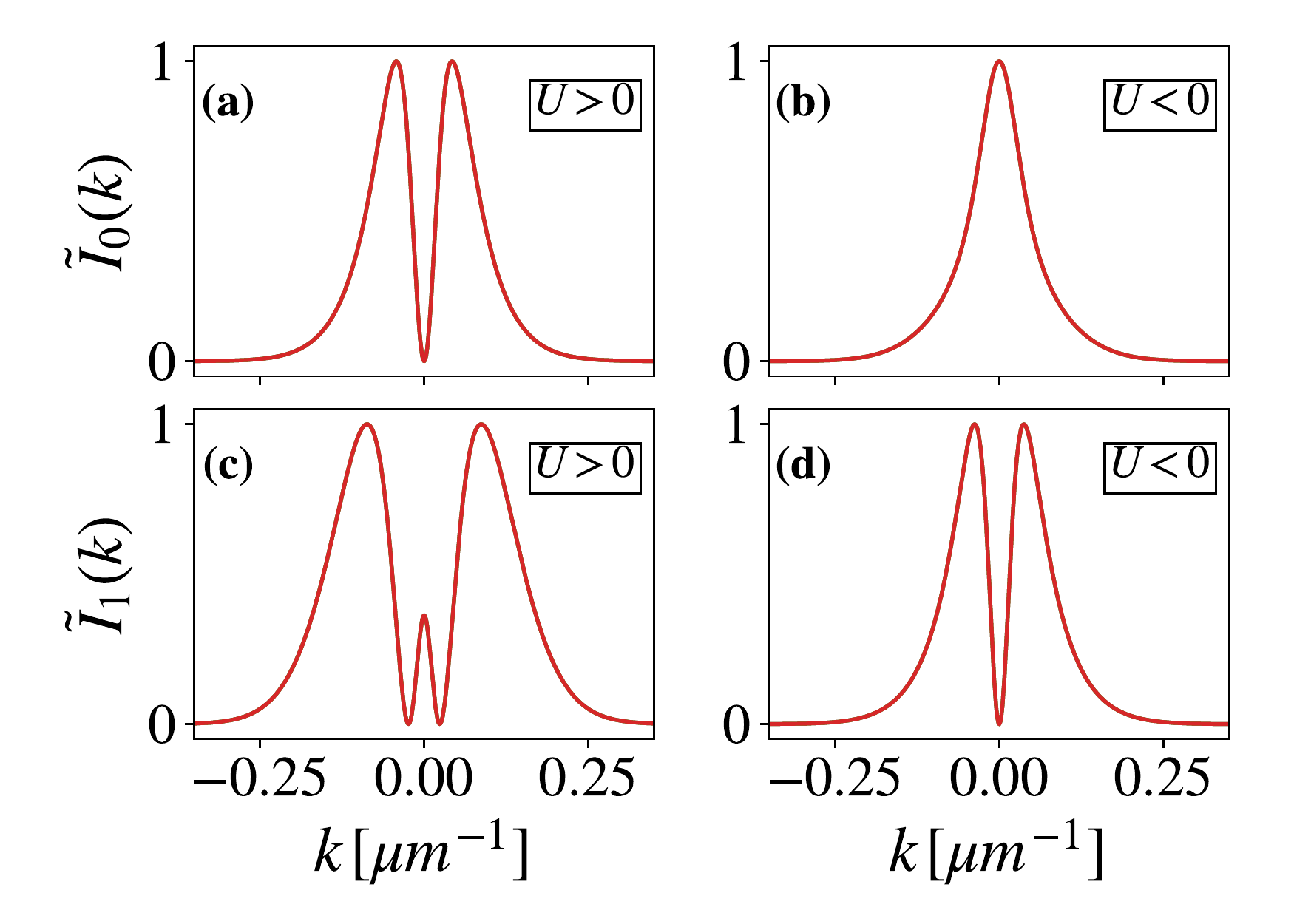}
\caption{Behavior of the spectral density $\tilde{I}_{n}(k)$ as a function of $k$ (in $\mu$m$^{-1}$), computed from the numerical solution for $\vec{\psi}_0(x)$ and (a) $\hbar U= 4.45$ meV and (b) $\hbar U=-4.45$ meV, as well as from $\vec{\psi}_1(x)$ for (c) $\hbar U= 4.45$ meV and (d) $\hbar U=-4.45$ meV. Data reported in all panels have been obtained for a repulsive barrier $V(x)$ with $V_0=6.25$ meV and $\sigma=10$ $\mu$m. All the other parameters are set as in Fig.~\ref{fig:spectral_density}. }\label{fig:numerical_emission_profiles}
\end{figure}

For completeness and clarity, we hereby report also the normalized spectral density associated to the bound-state solutions of the effective model, $\tilde{H}(x)$, for the same repulsive barrier used to obtain the out-of-equilibrium results reported in  Figs.~\ref{fig:spectral_density}(h) and (i), which is shown in Fig.~\ref{fig:numerical_emission_profiles}. In particular, in each panel of Fig.~\ref{fig:numerical_emission_profiles} we report the behavior of 
\begin{equation}
\tilde{I}_n(k)=\frac{I_n(k)}{\mathrm{max}\{I_n(k)\}},\quad I_n(k)=\left\vert A_{n,\,+}(k)+A_{n,\,-}(k)\right\vert^2,
\end{equation}
as a function of $k$. By direct comparison of Figs.~\ref{fig:numerical_emission_profiles}(a) and (c) to the numerical results shown in panel Fig.~\ref{fig:spectral_density}(h), we notice that there is a very good agreement between the spectral properties displayed by the output field for the two bound states, $\vec{\psi}_{0}(x)$ and $\vec{\psi}_{1}(x)$, and those displayed by the radiation emitted from the condensate. These results are also in remarkable good agreement with experimental outcomes obtained on similar devices \cite{Ardizzone_2022,Riminucci22,Gianfrate2023}. A similarly good agreement is displayed by the data obtained for $\hbar U=-4.45$ meV, and the corresponding ones in Fig.~\ref{fig:spectral_density}(i). We notice that the plot in Fig.~\ref{fig:numerical_emission_profiles}(d) corresponds to the second mode ($n=1$) supported by the potential, which is not displaying macroscopic occupation in this case.

\section{Summary and conclusion}\label{sec:summary}

We have reported an extensive theoretical analysis of the polariton condensation mechanism occurring in periodically patterned QW multilayers under incoherent driving. From a theoretical point of view, this work has required the generalization of the non-equilibrium Gross-Pitaevskii formulation under incoherent reservoir driving, originally proposed in Ref.~\cite{Wouters_Carusotto_PRL}, in order to account for multiple photonic bands interacting with the same QW exciton degenerate modes, which is an original contribution of the present work. 

After analyzing the complex eigenmodes of the non-Hermitian Hamiltonian model including exciton-photon coupling and losses, we have numerically solved the driven-dissipative dynamics in the presence of a continuous wave  Gaussian pump spot. We have modelled the effects of such a driving field as an effective potential barrier for the exciton-photon field, showing that it naturally induces the confinement of negative mass polaritons arising from the lower branch and becoming trapped due to the energy gap between the two polariton bands at energies below the exciton resonance.  

In addition, we have studied the emission characteristics of these polaritonic branches, which are associated to the sign of the diffractive coupling term introduced to mimic the effect of the periodic pattern on photonic eigenmodes. In particular, the positive sign of such diffractive coupling term is associated with the lower, negative mass branch being dark at normal incidence (BIC condition) with minimal losses, while if it is negative this same branch becomes bright at normal incidence and its imaginary part is much larger. Analysing the behavior of the model as a function of the pump intensity (effectively described by the height of this potential barrier), we have found that condensation actually occurs in such gap-confined eigenmodes, which become spatially quantized as a function of the depth of the confining potential. 
We then make the relevant conclusion that condensation in these systems is indeed the result of such gap-confinement of negative mass polaritons, irrespective of their dark or bright emission at normal incidence.
In particular, these results are in remarkable agreement with recent experimental findings corresponding to samples in which the diffractive coupling term is positive \cite{Ardizzone_2022,Riminucci22}. Hence, we predict that a similar phenomenology is going to occur even in the opposite case of a negative diffractive coupling term, which  motivates new experiments.

As a continuation of the present work, we envision applications of this theoretical framework to more complex pumping configurations, such as, e.g., periodic repetitions of a finite number of pumping spots, as recently reported in experiments \cite{Gianfrate2023}. In addition, a careful benchmarking of this effective model with experimental data might be a further test of the usefulness of the proposed approach. In particular, since experiments are often performed under pulsed excitations, an analysis of the solutions of the present model for pulsed driving, which is beyond the scope of the present manuscript, might be one of the possible next steps to undertake as a future project. We also notice that the model employed is a one-dimensional one, which essentially captures the relaxation mechanism in an effective way. On the other hand, the detailed dynamics along the transverse direction is completely neglected, which might still play a role in experiments. Hence, a possible further extension of the model might include the description of the modes dispersion along the transverse spatial direction as well.

\acknowledgments
We acknowledge useful discussions with L. C. Andreani, V. Ardizzone, A. Gianfrate, E. Maggiolini, H. C. Nguyen, H. S. Nguyen, F. Riminucci, D. Sanvitto, H. Sigurdsson, S. Zanotti.
This research was financially supported from the Italian Ministry of University and Scientific Research (MUR) through PRIN-2017 project ``Interacting Photons in Polariton Circuits'' (INPhoPOL), Grant No. 2017P9FJBS\_001.

\appendix

\section{Analytic expression of complex polariton dispeersion}\label{app:pol_bands}
As mentioned in Sec.~\ref{sec:polariton_bands}, the bands associated to the Hamiltonian model  $i\hbar \tilde{H}_0(k)$ (Eq.~\eqref{eq:model_no_potential}) can be determined analytically by looking for the four complex roots of the characteristic polynomial equation $P(\lambda)=\mbox{det}\left[E_0(k)-i\Gamma_0-\lambda \mathbb{1}_4 \right]$, in which ``$\mbox{det}$" denotes the matrix determinant. In what follows, we assume that $\sqrt{z}$ gives the square root a complex number $z$ with non-negative imaginary part, and we define 
\begin{equation}
E_X=\hbar(\omega_X-i\gamma_X),\,E_A=\hbar(\omega_A-i\gamma_A),\,\tilde{U}=U-i\tilde{\gamma}_A \, .
\end{equation}
Hence, the four eigenvalues of $i\hbar \tilde{H}_0(k)$ at each wave vector $k$, i.e., the polariton bands, are given by the following analytic expressions
\begin{equation}\label{eq:polariton_plus}
\begin{split}
&\lambda_{+,\,\pm}(k)=\frac{E_X+E_A-\sqrt{(\hbar \tilde{U})^2+(\hbar v_g k)^2}}{2}\\
&\pm\frac{1}{2}\sqrt{\left(E_X-E_A+\sqrt{(\hbar \tilde{U})^2+(\hbar v_g k)^2}\right)^2+4(\hbar\Omega_R)^2},
\end{split}
\end{equation}
and
\begin{equation}\label{eq:polariton_minus}
\begin{split}
&\lambda_{-,\,\pm}(k)=\frac{E_X+E_A+\sqrt{(\hbar \tilde{U})^2+(\hbar v_g k)^2}}{2}\\
&\pm\frac{1}{2}\sqrt{\left(E_X-E_A-\sqrt{(\hbar \tilde{U})^2+(\hbar v_g k)^2}\right)^2+4(\hbar\Omega_R)^2} \, ,
\end{split}
\end{equation}
respectively, corresponding to a total of four complex branches. In Eqs.~\eqref{eq:polariton_plus} and \eqref{eq:polariton_minus}, given the solution $\lambda_{\alpha,\,\beta}(k)$, the subscript $\alpha$ controls the sign in front of the $\sqrt{(\hbar U)^2+(\hbar v_g k)^2}$ term, while the subscript $\beta$ controls the sign in front of the square root containing the Rabi term, $4(\hbar\Omega_R)^2$.\\

\begin{figure}[t]
    \centering
    \includegraphics[scale=0.45]{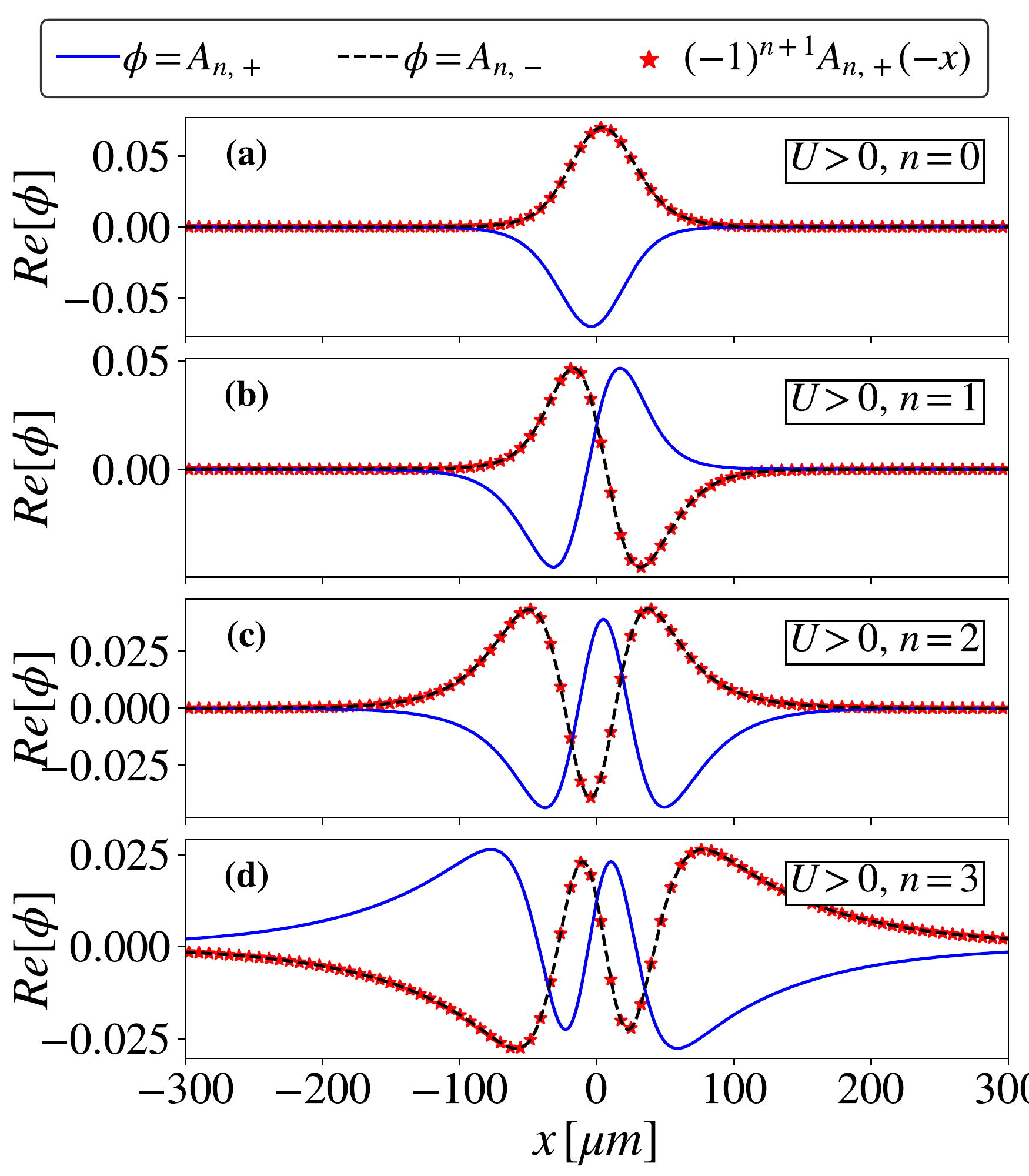}
    \caption{Spatially dependent real part of the first four eigenfunctions obtained from the model $\tilde{H}(x)$ for $V_0=2.5$ meV and $\sigma=35.0$ $\mu$m, assuming $\hbar U=+4.45$ meV and all the other relevant model parameters set as in Fig.~\ref{fig:example_eigenfunction0}.}
    \label{fig:components_large_well_BIC}
\end{figure}

\begin{figure}[t]
    \centering
    \includegraphics[scale=0.45]{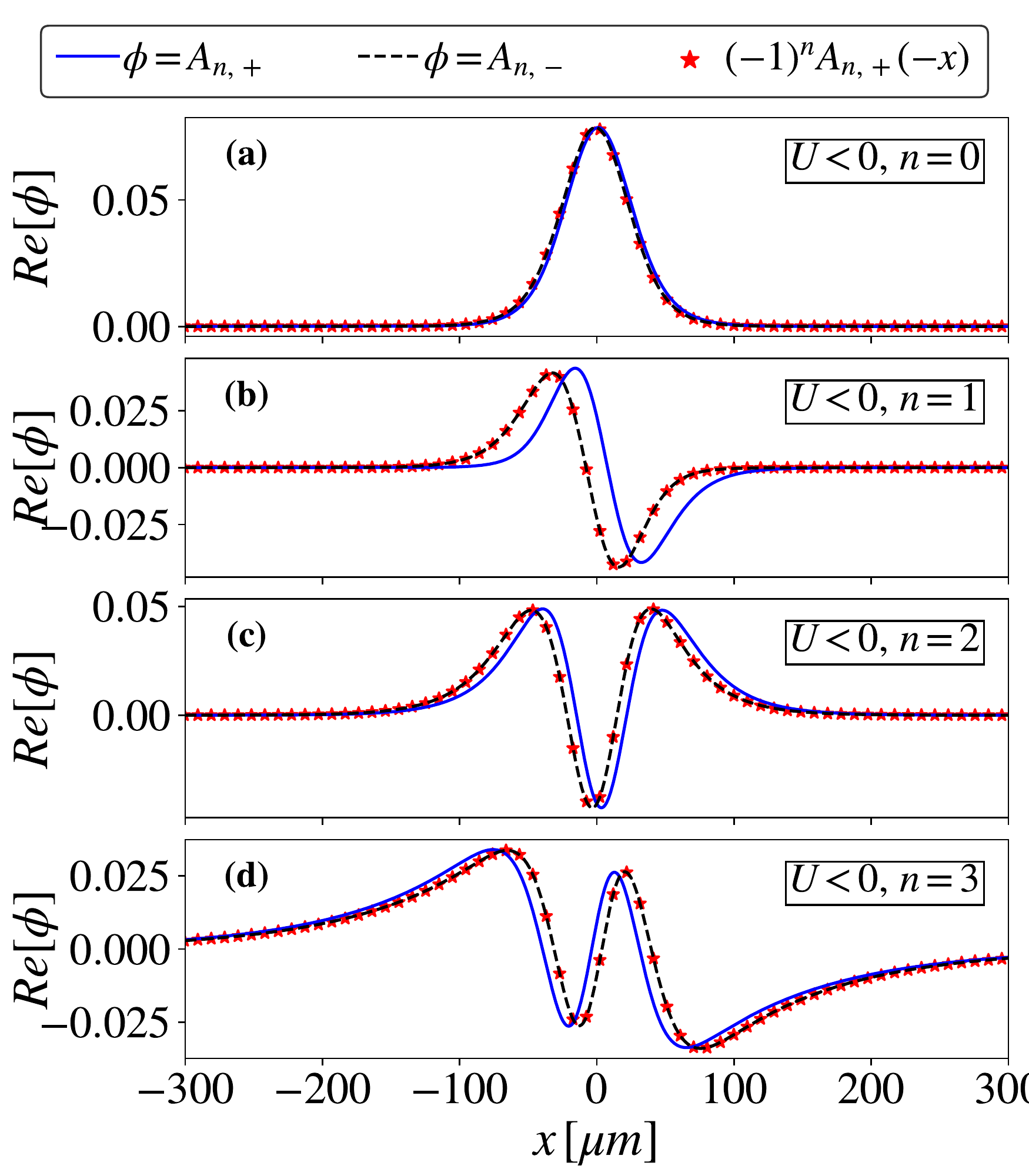}
    \caption{Spatially dependent real part of the first four eigenfunctions obtained from the model $\tilde{H}(x)$ for $V_0=2.5$ meV and $\sigma=35.0$ $\mu$m, assuming $\hbar U=-4.45$ meV and all the other relevant model parameters set as in Fig.~\ref{fig:example_eigenfunction0}.}
    \label{fig:components_large_well_LOSSY}
\end{figure}

\section{Symmetries and orthogonality of gap-confined states}\label{app:symm_gap_states}
We hereby report additional results concerning the inversion symmetry of the left- and right-moving photonic components, i.e., Eqs.~\eqref{eq:spatial_symmetry_Bic} and \eqref{eq:spatial_symmetry_Lossy} reported in the main text. In particular, here we report the real part of the photonic components of the first four eigenstates supported by a repulsive potential with $V_0=2.5$ meV and $\sigma=35$ $\mu$m, i.e., $\{\vec{\psi}_0,\,\vec{\psi}_1,\,\vec{\psi}_2,\,\vec{\psi}_3\}$. We first show the numerical results obtained for $\hbar U = +4.45$ meV in Fig.~\ref{fig:components_large_well_BIC}, while results obtained for $\hbar U=-4.45\,meV$ are reported in Fig.~\ref{fig:components_large_well_LOSSY}. Similar behaviors are obtained by considering the imaginary parts of the same components (not shown). As it is possible to see by direct comparison between the star-shaped markers and the dashed lines, for both positive and negative values of $U$ the numerical solutions are fully consistent with the two equations reported in Sec.~\ref{sec:gap_states}. In addition, due to the connection between the excitonic and photonic components reported in Eq.~\eqref{eq:excitonic_vs_photonic_components}, since $V(x)=V(-x)$, the exact same symmetry relations hold true also for $X_{\pm,n}(x)$. 

\begin{figure}[t]
    \centering
    \includegraphics[scale=0.6]{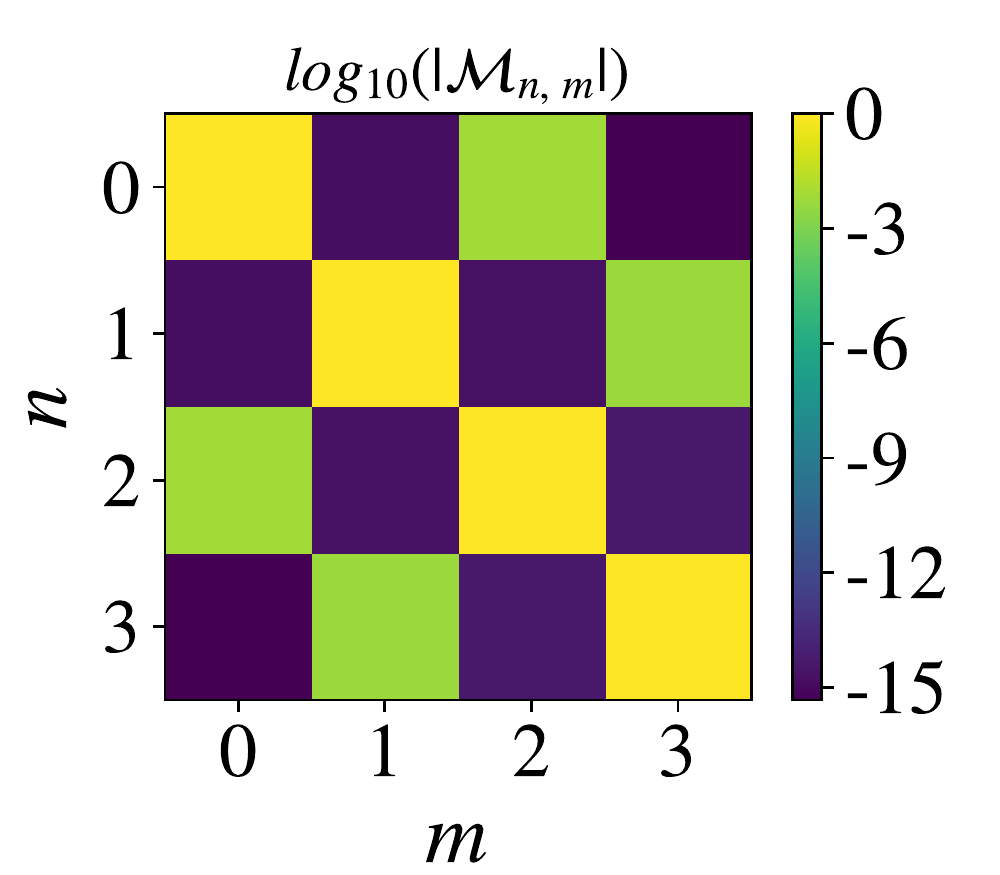}
    \caption{Absolute value of the overlap integral, $\mathcal{M}_{n,\,m}$, calculated for the first four eigenfunctions supported by $\tilde{H}(x)$ when $V_0=2.5$ meV and $\sigma=35.0$ $\mu$m, plotted in log scale (color bar on the side). Data have been obtained for $\hbar U=4.45$ meV. The other relevant parameters are set as in Fig.~\ref{fig:example_eigenfunction0}.}
    \label{fig:overlap_states}
\end{figure}

This is particularly relevant when considering the overlap integral between different eigenfunctions, i.e., formally defined as
\begin{equation}\label{eq:overlap_integral}
    \mathcal{M}_{n,\,m}\equiv\int \langle\vec{\psi}_{n}(x),\,\vec{\psi}_{m}(x) \rangle dx \, .
\end{equation}
Indeed, under the action of the inversion operation $x \to -x$, one obtains that the overlap integral transforms as follows
\begin{equation}
     \mathcal{M}_{n,\,m}\to (-1)^{m+n} \mathcal{M}_{n,\,m},
\end{equation}
which implies that $\mathcal{M}_{n,\,m}$ is identically zero whenever $n$ and $m$ have opposite parity. It would be tempting to conclude that, as in the Hermitian case, $\mathcal{M}_{n,\,m}=\delta_{n,m}$, which means that eigenfunctions corresponding to different eigenvalues are orthogonal. However, this does not seem be the case, as we have verified numerically. These results are summarized in Fig.~\ref{fig:overlap_states}, where we show the overlap integral \eqref{eq:overlap_integral} between the pairs of 4 eigenmodes shown in Fig.~\ref{fig:components_large_well_BIC}
is reported in log scale (see color bar).
While $\mathcal{M}_{n,\,m}$ is essentially  zero, up to numerical deviations, when $n$ and $m$ have opposite parity (in agreement with the discussion above), when $n$ and $m$ are different but have the same parity the overlap integral $\mathcal{M}_{n,\,m}\neq 0$. So in general we cannot conclude that eigenfunctions corresponding to different eigenvalues are always orthogonal.

\section{System initialization}\label{app:sys_init}
In this section we provide some details concerning the system initialization used to derive all the results shown in Sec.~\ref{sec:pol_cond_two}. In fact, in the time-evolution simulations the reservoir has been considered as initially empty, i.e., $n(x,t=0)=0$. For what concerns the exciton-photon subsystem, we consider as initial configuration a state where a tiny population is injected into the two bands $\lambda_{+,\,-}(k)$ and $\lambda_{-,\,-}(k)$. To do this, we took advantage of the explicit expression of polariton eigenstates in momentum space derived for $W(x)=0$. By denoting the eigenvector associated to $\lambda_{\alpha,\,\beta}(k)$ as $\vec{w}_{\alpha,\,\beta}(k)$, a generic system configuration where only the bands below the exciton resonance are populated can be expressed as:
\begin{equation}\label{eq:initial_configuration}
\vec{\psi}(x)=\int \frac{dk}{2\pi}e^{ikx}[c_{+,\,-}(k) \vec{w}_{+,\,-}(k)+c_{-,\,-}(k) \vec{w}_{-,\,-}(k)],
\end{equation}
where $c_{\alpha,\,\beta}(k)$ are complex coefficients. In the present analysis, we used a set of $\{c_{\alpha,\,\beta}(k)\}$ following a Gaussian distribution in $k$ space, namely
\begin{equation}
    c_{\alpha,\,\beta}(k)\propto \exp(-(k-k_0)^2/(2\Delta k)^2),
\end{equation}
with average $k_0$ and standard deviation $\Delta k$ drawn from two different uniform random distributions. Such sets of coefficients are then normalized in order to have an initial occupation corresponding to one particle, that is
\begin{equation}
\begin{split}
&N_{\psi}(t=0)=\int dx \langle \vec{\psi}(x),\,\vec{\psi}(x)\rangle=\\
&=\int \frac{dk}{2\pi}\left[\vert \tilde{c}_{+,\,-}(k)\vert ^2+\vert \tilde{c}_{-,\,-}(k)\vert^2\right] \, =1 .
\end{split}
\end{equation}
As already mentioned in the main text, the latter configuration is then evolved forward in time, and polariton condensation is monitored by looking at the time-behavior of the condensate occupation $N_{\psi}(t)$ for $t>0$.
%
\end{document}